\documentclass[amsmath,amssymb,aps,pra,superscriptaddress,onecolumn]{revtex4-2}

\usepackage{graphicx}

\begin{document}

\title{Low-frequency quantum sensing}

\author{E. D. Herbschleb}
\email{herbschleb@dia.kuicr.kyoto-u.ac.jp}
\affiliation{Institute for Chemical Research, Kyoto University, Gokasho, Uji-city, Kyoto 611-0011, Japan}
\author{I. Ohki}
\affiliation{Institute for Chemical Research, Kyoto University, Gokasho, Uji-city, Kyoto 611-0011, Japan}
\affiliation{Institute for Quantum Life Science, National Institutes for Quantum Science and Technology, Chiba 263-8555, Japan}
\author{K. Morita}
\affiliation{Institute for Chemical Research, Kyoto University, Gokasho, Uji-city, Kyoto 611-0011, Japan}
\author{Y. Yoshii}
\affiliation{Sumida Corporation, KDX Ginza East Building 7F, 3-7-2, Irifune, Chuo-ku, Tokyo, 104-0042, Japan}
\author{H. Kato}
\author{T. Makino}
\affiliation{National Institute of Advanced Industrial Science and Technology (AIST), Tsukuba, Ibaraki 305-8568, Japan}
\author{S. Yamasaki}
\affiliation{Kanazawa University, Kanazawa, Ishikawa 920-1192, Japan}
\author{N. Mizuochi}
\email{mizuochi@scl.kyoto-u.ac.jp}
\affiliation{Institute for Chemical Research, Kyoto University, Gokasho, Uji-city, Kyoto 611-0011, Japan}
\affiliation{Center for Spintronics Research Network, Kyoto University, Uji, Kyoto 611-0011, Japan}

\date{\today}

\begin{abstract}
Exquisite sensitivities are a prominent advantage of quantum sensors. Ramsey sequences allow precise measurement of direct current fields, while Hahn-echo-like sequences measure alternating current fields. However, the latter are restrained for use with high-frequency fields (above approximately $1$~kHz) due to finite coherence times, leaving less-sensitive noncoherent methods for the low-frequency range. In this paper, we propose to bridge the gap with a fitting-based algorithm with a frequency-independent sensitivity to coherently measure low-frequency fields. As the algorithm benefits from coherence-based measurements, its demonstration with a single nitrogen-vacancy center gives a sensitivity of $9.4$~nT~Hz$^{-0.5}$ for frequencies below about $0.6$~kHz down to near-constant fields. To inspect the potential in various scenarios, we apply the algorithm at a background field of tens of nTs, and we measure low-frequency signals via synchronization.
\end{abstract}

\maketitle

\section{Introduction}

Quantum sensing promises high-resolution sensors with unparalleled sensitivities by working with quantum properties such as coherence~\cite{Degen2017}. Nitrogen-vacancy (N-V) centers in diamond are high-potential candidates as such sensors for their extraordinary quantum mechanical properties even at room temperature~\cite{Rondin2014,Degen2017}, including long spin-coherence times~\cite{Herbschleb2019,Bar-Gill2013}. In conventional alternating current (ac) field-detection techniques with Hahn-echo and dynamical-decoupling schemes, the phase-accumulation time for the highest sensitivity is at around $T_2/2$~\cite{Herbschleb2019}, which dictates the lowest frequency measurable with high sensitivity. For frequencies far from $2T_2^{-1}$, the sensitivity becomes significantly worse. For higher frequencies, detection schemes have been proposed and demonstrated in the GHz range~\cite{Stark2017,Meinel2021}. On the other hand, the lowest frequency detected with a Hahn-echo sequence is $833$ Hz, as demonstrated with the longest $T_2$~\cite{Herbschleb2019}. Moreover, there is a significant amount of work focusing on direct current (dc) sensing with optically detected magnetic resonance (ODMR) measurements, which, although generally not specifically investigated, is envisaged to work for some low frequencies as well~\cite{Acosta2010a,Schoenfeld2011,Clevenson2015,Barry2016,Schloss2018,Zhang2021}.

Low-frequency sensing with high sensitivity is required for many applications. For example, it is useful for chemical structure analysis~\cite{Blanchard2013,Barskiy2019} and for searching particles beyond the standard model~\cite{Garcon2019,Wu2019} with low-field nuclear magnetic resonance (NMR) measurements. Contrary to NMR at high fields, at low fields, $J$ couplings, electron-mediated scalar couplings between spins in a molecule, are strongly represented. Since these are highly sensitive to the electronic structure of a molecule and its geometry, low-field spectra tend to be rather different for each molecule, while the differences in chemical shifts dominating at high fields could be small~\cite{Blanchard2013,Barskiy2019}. Moreover, since the inhomogeneous line width is proportional to the field strength, at low fields the line width and the signal-to-noise ratio improve significantly~\cite{McDermott2002}. Additionally, in conventional high-field NMR, resonant frequencies can be shifted down into the audiofrequency range (kHz and below), because this conversion enables filtering of high-frequency noise, and this is the region with high sensitivity for the phase detector~\cite{Hoult1975,Morris2017}.

Previously, low-frequency-like fields have been measured with continuous-wave (cw) ODMR techniques~\cite{Acosta2010a,Schoenfeld2011,Clevenson2015,Barry2016,Schloss2018,Zhang2021}. However, a drawback is the limited sensitivity compared to pulsed techniques, which becomes worse with longer coherence times~\cite{Dreau2011}. Alternatively, a pulsed-ODMR technique was proposed, which removes the laser-induced power broadening and as such improves the sensitivity significantly, although it is not as sensitive as coherence-based sequences still~\cite{Dreau2011}.

For sensing at zero and ultralow fields, recently a cw-ODMR technique was applied for an ensemble of N-V centers, measuring at a field below approximately $3$~\textmu T~\cite{Zheng2019}. The insensitivity normally expected for such techniques at low field was countered by applying circularly polarized microwave fields, which mostly affect one of each energy-level pair; the levels in each pair cross at zero field. In an alternative theoretical approach, a three-level system control was applied, which required a low bias field ($\le20$~G~\cite{Cerrillo2021}).

In the following, we present a fitting-based algorithm to measure low-frequency ac magnetic fields. With simulations and measurements, we explore the features of this algorithm; in principle, any low-frequency periodic field can be measured. We show that for low frequencies, which is below about $\left(2T_2^*\right)^{-1}$, the sensitivity of this algorithm is independent of frequency. Moreover, we employ the algorithm at a rather low background field to investigate the feasibility at such fields. Finally, we demonstrate the technique with synchronized low-frequency signals. Single N-V centers at room temperature are used for all experiments.

\section{Results}

\subsection{\label{Section:algorithm}Algorithm}
\begin{figure*}
	\includegraphics{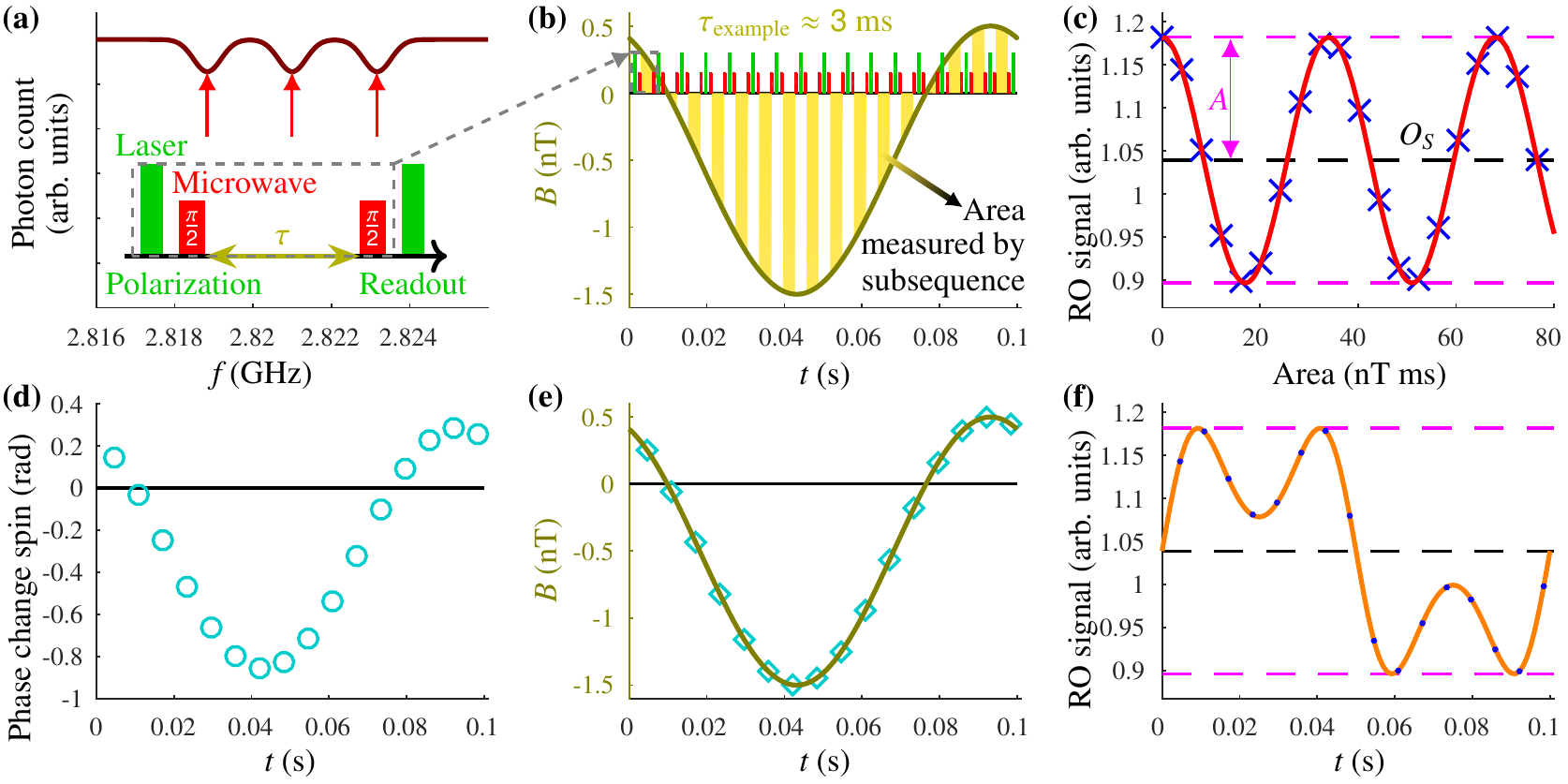}
	\caption{\label{Figure:algorithm}Algorithm. (a) Illustrative ODMR spectrum (dark red line at the top) with three valleys (indicated with red arrows) related to the energy differences between the $m_s=0$ and $m_s=-1$ states with hyperfine splitting due to the $^{14}$N nucleus. The inset shows the basic pulse sequence for FID measurements with delay $\tau$ between the MW $\pi/2$ pulses. (b) In the algorithm, a fixed-delay FID sequence is repeated during the period of the signal, measuring the yellow areas. Thus, the sequence functions as an oscilloscope with quantum measurements. (c) For a single sequence with the same fixed delay, by applying a dc field the spin rotates around the z axis depending on the size of the area (here the dc field magnitude times the fixed delay), resulting in an oscillation of the readout (RO) signal (simulation results with blue crosses, fit with red line). Such a measurement can be used as calibration, for example for the amplitude $A$ (magenta dashed lines are the extrema) and the offset $O_S$ (black dashed line) from Eq.~(\ref{Equation:fitting function}). (d) The final phase of the spin (analyzed simulation results with cyan circles) depends on the measured area during each subsequence. (e) By taking the shape of the areas into account while fitting, the parameters such as the ac field amplitude are retrieved (analyzed simulation results with cyan diamonds, resulting field with olive line). (f) With a single FID sequence the result would be ambiguous for larger fields given the rotational symmetry of the phase. However, the combination of multiple sequences allows some of this information to be retained (simulation results with blue dots, fit with orange line).}
\end{figure*}

The quantum measurement utilized in the algorithm is the free-induced decay (FID) sequence~\cite{Ramsey1950}, as displayed in the inset of Fig.~\ref{Figure:algorithm}(a). If the microwave (MW) frequency of the $\pi/2$ pulses is set exactly to the energy difference between the m$_s=0$ and m$_s=\pm1$ states, as appearing in ODMR spectra~\cite{Gruber1997} [Fig.~\ref{Figure:algorithm}(a)], the phase of the spin does not change during the time delay between the $\pi/2$ pulses. However, when a field is applied during this delay, the phase of the spin changes. For example, depending on the magnitude of an applied dc magnetic field, the spin rotates along the z axis, which results in an oscillation in readout signal [see, for example, Fig.~\ref{Figure:algorithm}(c)].

As illustrated in Fig.~\ref{Figure:algorithm}(b), in this algorithm, the sequence to measure a low-frequency ac field consists of repetitions of fixed-delay FID subsequences within the period of the field, which can be accumulated to obtain a sufficiently significant signal. Essentially, this is similar to a classical oscilloscope (with averaging), but with quantum measurements instead (which suggests the name ``QScope''), and the principle of repeating measurements is quite common, an example with N-V centers is with ODMR spectra~\cite{Schoenfeld2011}. For the resulting series of data points, each data point with readout signal $S$ follows from
\begin{equation}
\label{Equation:fitting function}
S = A\sin(\omega \int B\left(t\right)dt + \theta)+O_S,
\end{equation}
where $A$ is the amplitude, $\omega$ the frequency, $\theta$ the phase, and $O_S$ the offset of the oscillation in the signal, which stems from the magnetic field $B$ that rotates the spin. The parameters of this function are calibration constants (as explained in Supplemental Material~I) which follow from a calibration measurement giving a result as in Fig.~\ref{Figure:algorithm}(c), or they can be computed directly from the N-V center's parameters ($T_2^*$) and the time delay~\cite{Herbschleb2019}. For a sinusoidal ac field, the magnetic field $B$ at each time $t$ is given by
\begin{equation}
\label{Equation:ac field}
B = B_\textrm{ac}\sin(2\pi f t+\phi)+B_\textrm{dc},
\end{equation}
with $B_\textrm{ac}$ the field amplitude, $f$ its frequency, $\phi$ the phase, and $B_\textrm{dc}$ the constant field offset. Thus to find the ac field amplitude, this algorithm relies on fitting the data points of each subsequence to find the fitting parameter $B_\textrm{ac}$.

However, the field at each data point is not found directly, as the readout signal only gives the final phase of the spin at the end of each subsequence, as plotted in Fig.~\ref{Figure:algorithm}(d). To retrieve the measured field accurately, the shape of the field during the subsequence needs to be taken into account, as implied by the integral in Eq.~(\ref{Equation:fitting function}). By fitting the readout signal directly utilizing integrals, the field is retrieved accurately [Fig.~\ref{Figure:algorithm}(b,e)].

In Fig.~\ref{Figure:algorithm}(f), a directly fitted readout signal is plotted. This illustrates an additional advantage of the algorithm, which is an inherent increase in dynamic range for ac fields. This is a consequence of the relatively slow increase of field over time, which allows us to determine how often the spin rotated fully by $2\pi$ at the extrema of the ac field. Generally, multiple measurements have the ability to increase the dynamic range~\cite{Herbschleb2021}.

Any periodic function can be fitted, though one requirement out of two is necessary to perform the measurement: either the period of the signal needs to be known, or a way of synchronization must exist, for example via triggered measurements. The remainder of the parameters results from the measurement, for example, in Fig.~\ref{Figure:algorithm}(e) the phase and dc component of the sinusoid are found as well. Throughout this paper, the main focus is on measuring low-frequency ac fields. In other words, we measure the amplitude, the result of which is independent of parameters such as the phase and the dc component.

\subsection{Sensitivity definition}
The sensitivity of a measurement is its uncertainty times the square root of the measurement time~\cite{Taylor2008,Herbschleb2019}. Therefore, for this fitting-based algorithm, the sensitivity $\eta_\textrm{coef}$ of each fitting coefficient $\textrm{coef}$ is
\begin{equation}
\label{Equation:sensitivity}
\eta_\textrm{coef} = \sigma_\textrm{coef} \sqrt{T_\textrm{meas}},
\end{equation}
with $\sigma_\textrm{coef}$ the uncertainty of the respective fitting coefficient, and $T_\textrm{meas}$ the measurement time. The $\sigma_\textrm{coef}$ follows from fitting the measurement data, which allows computation of the standard error (uncertainty) of the fitting coefficients.

The sensitivity depends on the time delay between the $\pi/2$ pulses in the sequence. The optimum is derived in Supplemental Material~I \footnote{See Supplemental Material for derivations and supportive information.}. At first, the linear regime is investigated, since the sensitivities given by Hahn-echo measurements are based on a single point (at the maximum gradient), which is the linear regime. Moreover, the sensitivity of ODMR techniques is based on the maximum gradient of a valley in the ODMR spectrum [Fig.~\ref{Figure:algorithm}(a)], which is the linear regime as well. Therefore, this allows a fair comparison with these standard methods.

\subsection{Measurement}
The sample measured throughout this paper consists of $n$-type diamond. It is epitaxially grown onto a Ib-type (111)-oriented diamond substrate by microwave plasma-assisted chemical-vapor deposition with enriched $^{12}$C ($99.998\%$) and with a phosphorus concentration of approximately $5\times10^{16}$ atoms cm$^{-3}$~\cite{Kato2016,Herbschleb2019}. We target individual electron spins residing in single N-V centers with a standard in-house built confocal microscope; each set of experiments works with a different N-V center, as locations are lost in between experiments. MW pulses are applied via a thin copper wire. Since the nitrogen atom causes hyperfine splitting of the $m_s=\pm1$ states, to ensure maximum contrast in our measurements, each energy difference is addressed with a separate MW source [so imagine all three frequencies indicated by arrows in Fig.~\ref{Figure:algorithm}(a) are applied], unless stated differently. Magnetic fields are induced with a coil near the sample. All experiments are conducted at room temperature.

\begin{figure*}[b]
	\centering
	\includegraphics{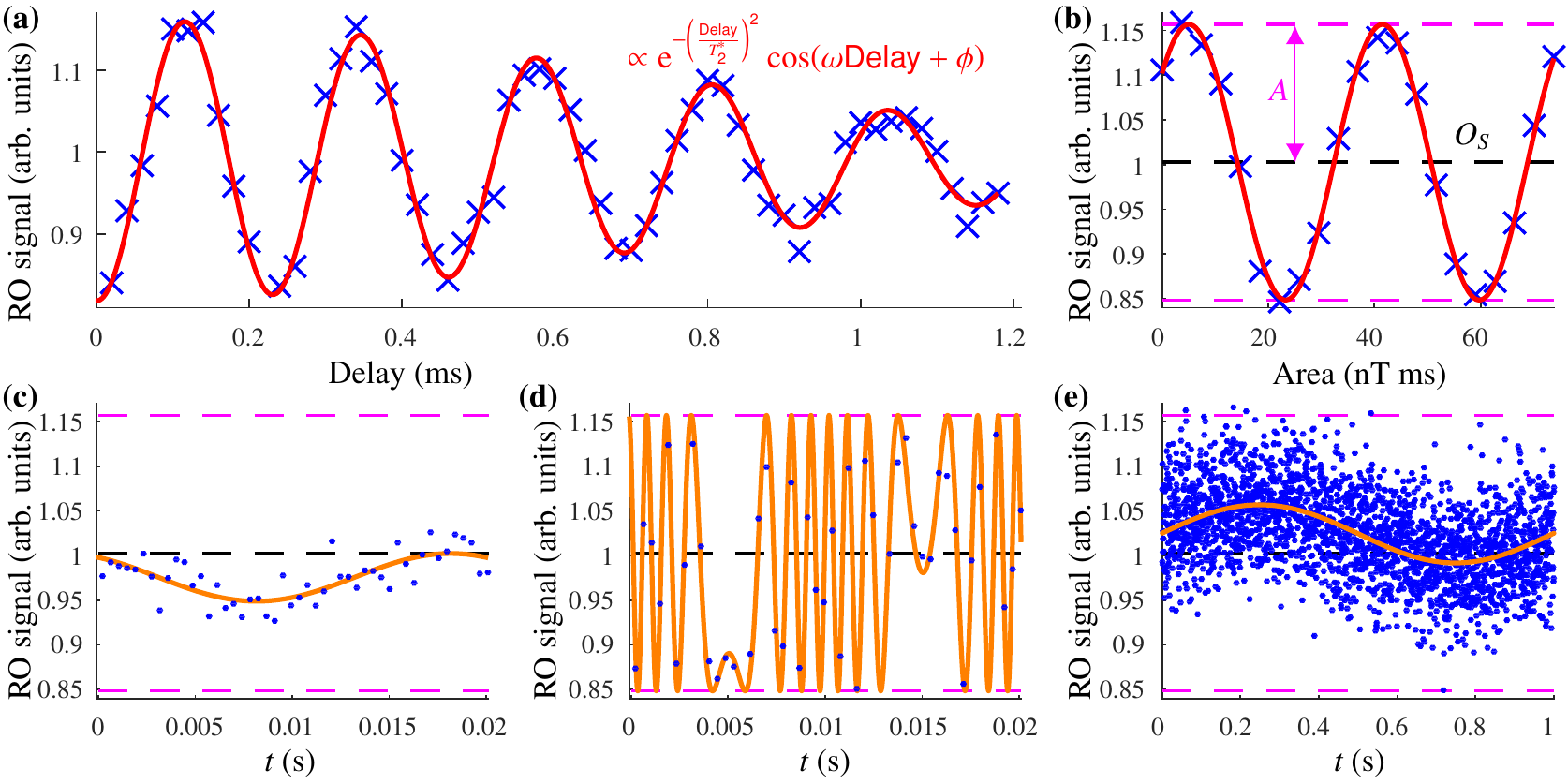}
	\caption{Exploratory measurements. (a) Readout (RO) signal versus delay giving the FID measurement result (data with blue crosses, fit with red line) to determine $T_2^*$ ($1.05_{-0.05}^{+0.05}$~ms). (b) Calibration measurement for a delay of $0.4$~ms (data with blue crosses, fit with red line). The horizontal black dashed line indicates the offset $O_S$, while the horizontal magenta dashed lines give the extrema due to amplitude $A$ [see Eq.~(\ref{Equation:fitting function})]. (c) Example measurement exhibiting the independence of phase and offset when measuring an ac field ($50$-Hz signal with an amplitude of $3.1$~nT, data with blue dots, fit with orange line). (d) Example measurement for a high dynamic-range measurement, which is outside the linear regime (data with blue dots, fit with orange line). The amplitude is $100$ times the amplitude in (c) and (e). (e) Example measurement for a field of $1$~Hz with an amplitude of $3.1$~nT (data with blue dots, fit with orange line). The offset field of the N-V center for the measurements in this figure is about $1.8$~mT.}
	\label{Figure:measurement}
\end{figure*}

N-V centers with longer inhomogeneous dephasing times ($T_2^*$) are more beneficial for the sensitivity of coherence-based sensors, therefore, a center with a $T_2^*$ of about $1$~ms is chosen for the initial experiments [see Fig.~\ref{Figure:measurement}(a)]. Although the optimum time delay is about $T_2^*/2$ (Supplemental Material~I), a slightly shorter time delay is applied. The reason is that while the sensitivity is not significantly worse for delays near the optimum, the apparent $T_2^*$ of a measurement decreases with measurement time due to environmental effects~\cite{Herbschleb2019}. So even though the actual $T_2^*$ of the N-V center is longer than $1$~ms (see Supplemental Material~II), since long measurements are required to accurately estimate the sensitivities for low frequencies, the apparent $T_2^*$ might be less for these measurements. Hence, we choose a time delay of $0.4$~ms, since it ensures a sensitivity within $10\%$ of the optimum for the considered range of apparent $T_2^*$s (see Supplemental Fig.~2).

First, the algorithm principles described in the previous section are elucidated with example measurements. The calibration data, equivalent to Fig.~\ref{Figure:algorithm}(c), are depicted in Fig.~\ref{Figure:measurement}(b). The contrast with a time delay of $0.4$~ms is close to $30\%$, as expected with $T_2^*\approx 1$~ms. In Fig.~\ref{Figure:measurement}(c), a sinusoid with nonzero phase and dc component is measured to demonstrate the independence of such parameters for getting the ac amplitude, here $3.1$~nT. The measurement of a signal with an amplitude beyond the standard dynamic range (of a single-sequence measurement), shown in Fig.~\ref{Figure:measurement}(d), illustrates the increase of the dynamic range by measuring a signal with an amplitude of $0.31$~\textmu T. Finally, Fig.~\ref{Figure:measurement}(e) plots a measurement result for the lowest frequency measured ($1$~Hz) with an amplitude of $3.1$~nT. This visualizes that a large number of data points, although with significant individual spread, indeed resemble accurately the ac field.

\subsection{Sensitivity measurement}
To inspect the performance of the algorithm, the sensitivity is calculated and measured for a range of frequencies. Since the measurement time [see Eq.~(\ref{Equation:sensitivity})] follows from the period of the signal (period duration times number of accumulations), this is a proper figure of merit, which includes all overhead time. Moreover, it allows comparison with the standard Hahn-echo measurement. For the latter, to look at its best possible sensitivity, we ignore its overhead time. However, the disadvantage in the comparison for our algorithm is rather small, since at low frequencies, the overhead time of the standard Hahn-echo measurement would be negligible. As additional comparisons, we compute the theoretically best pulsed-ODMR sensitivity for the measured N-V center~\cite{Taylor2008,Dreau2011}, the averaged sensitivity over low frequencies for the longer measurement times by fitting a Lorentzian in the Fourier spectrum~\cite{Glenn2018}, and the dc-field sensitivity of the cw-ODMR technique for our sample~\cite{Dreau2011}.

\begin{figure}[b]
	\centering
	\includegraphics{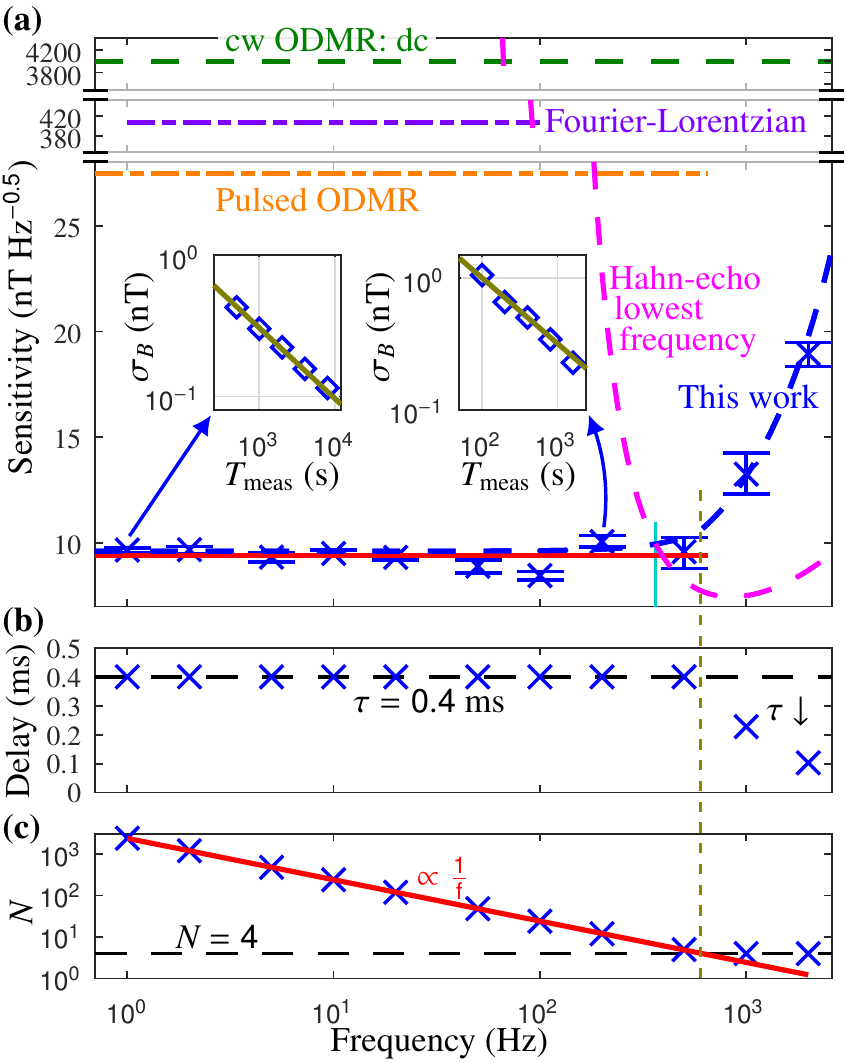}
	\caption{Sensitivity. (a) Sensitivity versus frequency: data points with blue crosses (error bars indicate standard errors); calculated result with blue dashed line; fitted result for low frequency with horizontal red line ($\eta_\textrm{ac}=9.4_{-0.2}^{+0.1}$~nT~Hz$^{-0.5}$); calculated result with Hahn-echo sequence for the current longest $T_2$ with magenta dashed line (adjusted from Ref.~\cite{Herbschleb2019} with the parameters of the current experimental setup for fair comparison); calculated result for optimal pulsed ODMR~\cite{Dreau2011} with orange dashed-dotted line; averaged result over low frequencies for Fourier-Lorentzian as a guide to the eye with purple dashed-dotted line; calculated result for dc sensing via cw ODMR as a guide to the eye with green dashed line at the top (for the current N-V center and experimental setup, adjusted from Ref.~\cite{Dreau2011}); threshold frequency with vertical cyan line ($f_\textrm{threshold}=0.36$~kHz). Insets show two examples of the amplitude uncertainty versus measurement time $T_{meas}$ results (data with blue diamonds, fits with olive lines), where fits give the sensitivities (left for $1$~Hz and right for $0.2$~kHz). Note the breaks on the vertical axis. Here, the offset field of the N-V center is about $1.8$~mT. (b) The fixed delay applied in the subsequences for each frequency (blue crosses), with a maximum of $0.4$~ms (horizontal black dashed line). For high frequencies, this delay decreases since a minimum number of data points are required per period. (c) Number of subsequences $N$ (blue crosses, fit with red line) per period, thus data points, for each frequency, with a minimum of $4$ (horizontal black dashed line). The vertical olive dashed line is a guide to the eye at the maximum frequency where $N>4$.}
	\label{Figure:sensitivity}
\end{figure}

The calculated sensitivity, explained in more detail in Supplemental Materials~I~and~III, is drawn in Fig.~\ref{Figure:sensitivity}(a). This shows that the sensitivity is expected to be frequency independent below a certain threshold frequency (see Supplemental Material~III). This is sensible, since when halving the frequency, the period and hence measurement time doubles, but the number of data points $N$ also doubles. Since $\eta\propto T_\textrm{meas}^{0.5}$ and $\eta\propto\sigma_\textrm{coef}\propto N^{-0.5}$, the sensitivity is constant. Above the threshold, the sensitivity becomes worse simply because at least several data points are required in one period of the signal [see Fig.~\ref{Figure:sensitivity}(c)]; four are chosen here for fitting the four unknowns of the current signal shape. Compared to Nyquist's sampling theorem, which states that the signal must be sampled at a rate over $2$ times the highest frequency component in order to reconstruct it faithfully, a higher sampling rate is required, since we look at a finite time of a single period only. Either way, it is reflected by the measurements at higher frequencies, where the time delay between the MW pulses decreases linearly to maintain a sufficient number of points [with the period of the signal, see Fig.~\ref{Figure:sensitivity}(b)], and hence the sensitivity worsens (roughly proportional to $T_\textrm{period}^{-0.5}$, until the overhead time becomes significant). Moreover, for the highest two frequencies (which are outside the studied low-frequency regime), multiple periods are measured, as fitting four unknowns on four data points is often mathematically possible, but having more data points than parameters is preferred for fitting.

The uncertainty $\sigma_\textrm{ac}$ is measured for a number of measurement times for frequencies ranging from $1$~Hz to $2$~kHz, and the results are fitted to Eq.~(\ref{Equation:sensitivity}) to determine the sensitivity for each frequency [see insets of Fig.~\ref{Figure:sensitivity}(a)]. The results are added to Fig.~\ref{Figure:sensitivity}(a); they are consistent with the calculated results. The low-frequency sensitivity is $9.4_{-0.2}^{+0.1}$ nT Hz$^{-0.5}$. For lower frequencies, the sensitivities become slightly worse, which we attribute to the earlier mentioned decay of the apparent $T_2^*$, since determining the sensitivities for these frequencies takes more time. The results and explanations for the other fitting parameters are given in Supplemental Material~IV.

\subsection{Low-field measurement}
\begin{figure}
	\centering
	\includegraphics{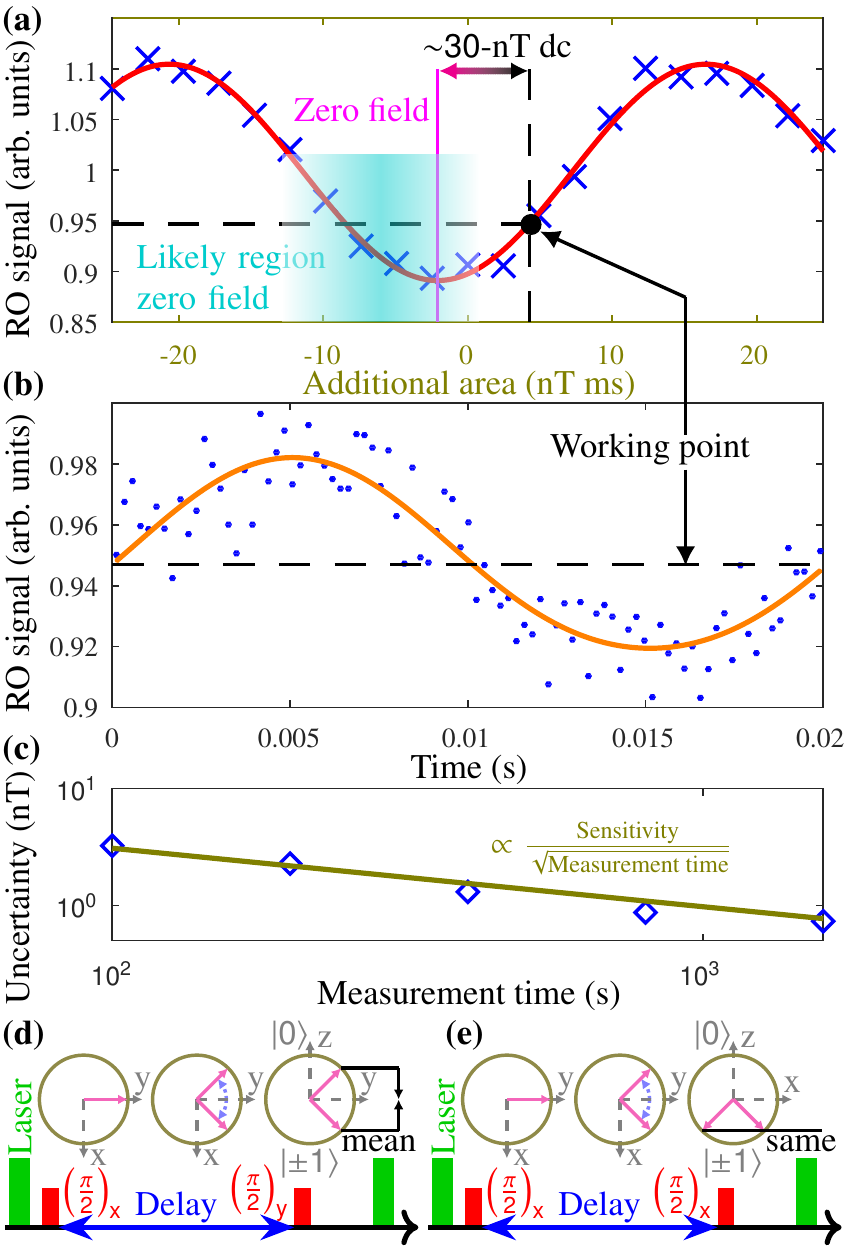}
	\caption{Low-frequency sensing at low field. (a) Calibration measurement for a delay of $0.2$~ms around zero field (RO signal is readout signal, data with blue crosses, fit with red line). The cyan-shaded area indicates the range of fields that include the actual zero field based on FID measurements (Supplemental Material~V), while the extremum (vertical magenta line) in this region is when the additional field cancels the remaining background field exactly. The measurements in (b) and (c) are performed at a background field of $30$~nT (black dashed lines). (b) Example measurement for a field of $50$~Hz with an amplitude of $10$~nT at a background field of $30$~nT (data with blue dots, fit with orange line). (c) Uncertainty versus measurement time to find the sensitivity for the same signal as in (b) (data with blue diamonds, fit with olive line). (d) Near zero field, when the preparation MW pulse (first lower red rectangle after the laser pulse), here $\left(\pi/2\right)_x$, and the readout MW pulse, here $\left(\pi/2\right)_y$, are along different axes, the result depends on the spin state and roughly averages towards a field-independent value (black arrows). (e) When the readout MW pulse is along the same axis, here $\left(\pi/2\right)_x$, the readout result is the same (black line) for each spin state, enabling measurement of the field.}
	\label{Figure:Low field}
\end{figure}

Measuring at low field adds the complexity of level (anti)crossings, which could render a sensor insensitive. Therefore, to investigate the algorithm at low fields, first, we cancel the field in the $z$ direction to below approximately $1$~nT, as explained in detail in Supplemental Material~V, which results in overlapping energy levels of the positive and negative spin states. For the sensing experiment, instead of the previous three MWs, we use a single MW for the lower-frequency transition only. We set the frequency of this MW to the energy difference at zero field. Now, if any magnetic field is applied, the two overlapping energy levels change equally in opposite directions. Thus, as long as the final MW pulse in each FID sequence is along the same axis as the preparation MW pulse, both possible states have the same effect on the readout signal, even though their spins rotate in opposite directions effectively [see Fig.~\ref{Figure:Low field}(d,e)].

To measure a low-frequency field, the background field needs to remain sufficiently constant during the measurement; for example, the daily fluctuation in the earth magnetic field is in the order of tens of nTs~\cite{Hitchman1998}. To limit the measurement time, we choose a frequency of $50$~Hz and we use a delay of $0.2$~ms. The calibration measurement is displayed in Fig.~\ref{Figure:Low field}(a), for which the background field is close to $0$~T. FID measurements locate the zero field within a range of tens of nTs, such that the valley in the calibration measurement in this range gives the actual zero field, here with a precision of $0.7$~nT (see Supplemental Material~V).

For the low-frequency measurement, instead of measuring at exact zero, we measure closer to the linear regime, resulting in a background field of $\sim$$30$~nT. An example measurement is plotted in Fig.~\ref{Figure:Low field}(b); the slight asymmetry around the center horizontal indicates that we are at the edge of the linear regime. At this background field, we measure the sensitivity, which is $31_{-1}^{+1}$~nT~Hz$^{-0.5}$ [Fig.~\ref{Figure:Low field}(c)]. This low-field sensitivity is worse compared to the high-field sensitivity, mostly owing to the $2$ times shorter time delay and the single-tone MW (instead of multitone), and to a lesser degree owing to the lower coherence time of the measured N-V center, the vicinity of the nonlinear regime, and a nonperfect detuning (Supplemental Material~V).

\subsection{Synchronized measurement}
So far we study signals with a known period, but as mentioned in Section~\ref{Section:algorithm}, an alternative is measuring synchronized signals, such as NMR signals. To obtain an example low-frequency NMR signal, the sample (water or ethanol) is placed inside a permanent magnet (approximately $1$~T) at room temperature, and with a radio-frequency (rf) $\pi/2$ pulse emitted via a coil around the sample, its nuclear spins are excited. This pulse marks the synchronization time. The transient emitted field, so here its free nuclear precession response, is picked up via the same coil, which is connected to the coil around the N-V center via a switch in order to record it by the N-V center. Owing to mixing the emitted signal with a reference oscillator, a down-converted low-frequency signal is obtained (see the Appendix for details). Since the signals decay and thus a large difference in amplitude between the start and the end exists, a delay of $0.1$~ms is chosen to allow higher amplitudes to fit the linear regime, while also easing the environmental effects given the required long measurements~\cite{Herbschleb2019}, reducing the sensitivity by less than $2$ times compared to the initial measurements (Figs.~\ref{Figure:measurement}~and~\ref{Figure:sensitivity}).

\begin{figure}
	\centering
	\includegraphics{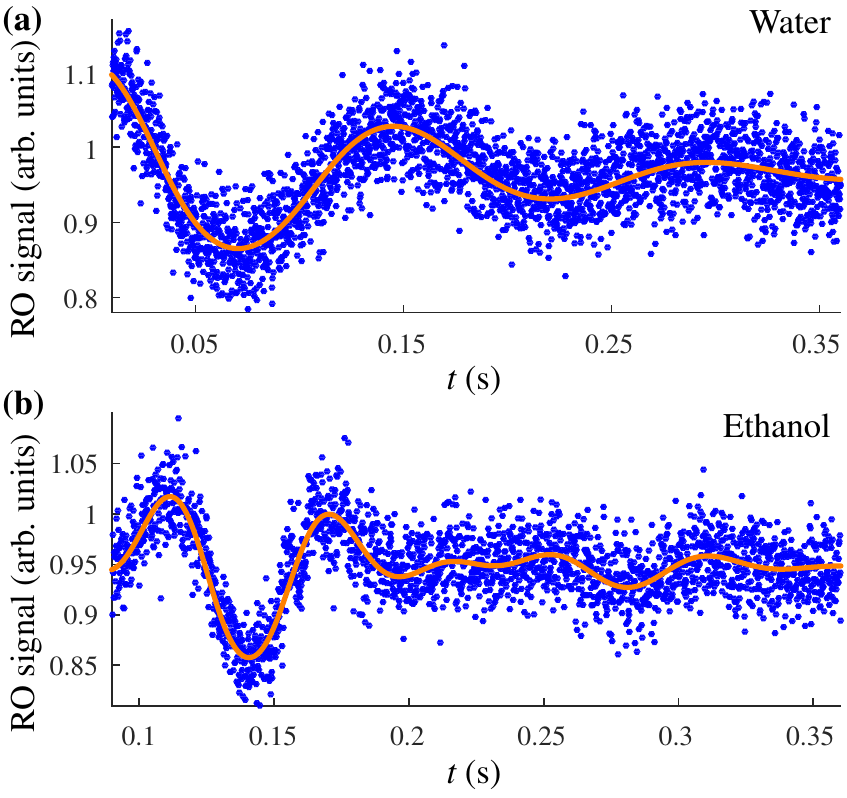}
	\caption{Synchronized low-frequency sensing. (a) Free nuclear precession readout (RO) signal versus time for part of the time axis (total length is $1.0$~s, repeated $10000$~times, data with blue dots, fit with orange line) measuring the response of a deionized water sample after applying a rf $\pi/2$ pulse. Here, the amplitude of the low-frequency signal ($7$~Hz) starts at $43.7$~nT (left side), and decays to $2.5$~nT (right side). (b) Free nuclear precession readout signal versus time for part of the time axis (total length is $1.0$~s, repeated $20000$~times, data with blue dots, fit with orange line) measuring an ethanol sample after applying a rf $\pi/2$ pulse. Here, the amplitudes of the low-frequency signal start at $9.6$~nT ($7$~Hz), $19.7$~nT ($14$~Hz), and $13.6$~nT ($21$~Hz), and decay to $0.4$~nT, $0.9$~nT, and $0.6$~nT, respectively. The offset field of the N-V center for the measurements in this figure is about $0.5$~mT.}
	\label{Figure:NMR}
\end{figure}

For the straightforward case of deionized water, the result is plotted in Fig.~\ref{Figure:NMR}(a). A low-frequency signal of about $7$~Hz is measured, the amplitude of this signal decays from $43.7$ to $2.5$~nT in $0.35$~s due to the inhomogeneous dephasing time $T_2^*\approx0.2$~s. On the other hand, ethanol has a more involved spectrum given the three discernible proton groups, of which notably the CH$_3$ and CH$_2$ groups are sufficiently close to have split peaks caused by $J$ coupling. As the focus is on low-frequency measurements, we aim to measure the peaks of the CH$_3$ group which are shifted to around $14$~Hz. The readout signal is shown in Fig.~\ref{Figure:NMR}(b), and the amplitudes of the three low-frequency signals ($7$, $14$, and $21$~Hz) are $9.6$, $19.7$, and $13.6$~nT initially, and decay to $0.4$, $0.9$, and $0.6$~nT in $0.27$~s with $T_2^*\approx0.2$~s. The roughly $1$:$2$:$1$ structure in amplitude with a peak-to-peak difference of $7$~Hz is in good agreement with the known $J$-coupling value of ethanol’s CH$_3$ group ($^3J_{\textrm{H},\textrm{H}}=6.9$~Hz~\cite{Bovey1967}).

\section{Discussion}

To compare our results with the standard coherent method for measuring ac fields, the Hahn-echo sequence~\cite{Hahn1950a}, we use the results of the single N-V center with the longest coherence time published so far~\cite{Herbschleb2019}, as this allows measurement of the lowest frequency. The frequency dependency of this sensitivity is added to Fig.~\ref{Figure:sensitivity}(a). Naturally, since our fitting-based algorithm needs at least several data points each period while the Hahn-echo measurement requires just one, our sensitivity at higher frequencies is worse compared to the Hahn-echo measurement. However, the Hahn-echo sensitivity at low frequencies becomes exponentially worse due to a finite $T_2$ ($2.4$~ms). The threshold frequency is about $0.36$~kHz.

As an alternative, we analyze the measurement data by applying the Fourier transform and fitting a Lorentzian to find the amplitudes, which is a common way to process the frequency spectrum~\cite{Glenn2018}. As this is rather sensitive to the noise near the single sharp peak (see Supplemental Material~VI), the uncertainties vary greatly; the example in Supplemental Material~VI is the most common case. Hence, to get an impression of the low-frequency amplitude sensitivity, the mean of the sensitivities for the long measurements times (thus the noise is averaged for longer) is added to Fig.~\ref{Figure:sensitivity}(a). Although the sensitivity is significantly worse compared to the one from the time-domain analysis, care should be taken when interpreting these results. Depending on the application, processing the data in the frequency domain with different methods (instead of Lorentzian fitting) could be suitable and could result in improved sensitivities. Nonetheless, when looking at the spectral resolution, time-domain fitting methods, such as harmonic inversion~\cite{Mandelshtam1997}, which finds the frequency and amplitude of $K$ combined cosines given $N$ data points, have an in principle ``infinite'' frequency resolution, thus the Fourier transform seems lacking with a $T^{-1}$ spectral resolution. Once again, since both domains do contain all information, it might be possible to extract the same spectral information from the frequency domain as well, but it seems that the time domain has the more convenient methods.

The dc sensitivity for our N-V center utilizing the well-known cw-ODMR method~\cite{Dreau2011} is drawn in Fig.~\ref{Figure:sensitivity}(a) as well. It shows that the dc sensitivity is over $2$ orders of magnitude worse; any low-frequency algorithm would have a sensitivity strictly worse than the dc sensitivity. This improvement is as expected, since coherence-based methods are more sensitive and benefit more from longer coherence times than noncoherence-based ones~\cite{Dreau2011}. A reason is the power broadening due to the laser and the MW. When comparing with recent results of cw-ODMR experiments, the sensitivity of our algorithm with a single N-V center is comparable to the sensitivity of earlier ODMR techniques with ensembles (for example with $10^{13}$ N-V centers giving $\eta\approx2.9$~nT~Hz$^{-0.5}$~\cite{Clevenson2015}). For fair comparison, for both techniques all overhead is included. With the single N-V center in the current experiment, a rather high spacial resolution is possible compared to ensembles. More recent ODMR techniques with ensembles have an improved sensitivity, however the lowest frequency is about $5$~Hz~\cite{Schloss2018}, while our algorithm has no lower bound in principle. Additional technical improvements have enhanced the sensitivity even further, as for example with flux concentrators for a bandwidth of $20$-$200$~Hz~\cite{Zhang2021}.

The power broadening induced by the laser can be removed by utilizing the pulsed-ODMR method. However, the sensitivity is worse by a factor of $\sqrt{2e}$, with $e$ Euler's number, and an additional reduction occurs due to a loss in contrast~\cite{Dreau2011}. This theoretical sensitivity, based on the results with the Ramsey sequence (thus assuming fitting as well), is plotted in Fig.~\ref{Figure:sensitivity}(a). We are not able to verify this theoretical best sensitivity for pulsed ODMR, as the contrast decreased rapidly while elongating the MW pulse. The pulsed ODMR in Supplemental Material~V utilizes a MW pulse length of $1.5$~\textmu s, which is still orders of magnitude away from the optimum around $T_2^*$~\cite{Dreau2011}. Likely, this is a technical limitation only. Nonetheless, compared to our results with the Ramsey sequence, the sensitivity is worse. Moreover, the range of measurable field amplitudes is significantly lower, since this is limited by the line width of the pulsed-ODMR spectrum, hence a measurement such as in Fig.~\ref{Figure:measurement}(d) would not be possible.

Besides measuring the ac amplitude of a field, the dc component, the phase and potentially the frequency (for measurements using synchronization only) follow as well (see Supplemental Material~IV). For the dc component, although it follows from fitting, the sensitivity is the same as expected from its standard method (a single fixed-delay FID sequence~\cite{Taylor2008}). This is intuitive, since within the same measurement time, the sequence is repeated the same number of times for both. Moreover, for this algorithm, the ac sensitivity shows to be about $\sqrt{2}$ worse than the dc sensitivity, as is expected from theory (see Supplemental Material~I). The intuition is that for fitting a constant, a single data point suffices, while fitting an amplitude, which essentially is a difference between two points, requires double the points. Since, as mentioned before, $\eta\propto N^{-0.5}$, this gives the factor of $\sqrt{2}$.

Although the linear regime of the algorithm is investigated so far to compare with other methods, it can be extended to work outside this regime. This allows measurement with a higher dynamic range, as, for example, shown in Fig.~\ref{Figure:measurement}(d). Multiple readout phases are required (this is possible via changing the phase of the MW pulses), and the resulting sensitivity is $\sqrt{2}$ worse compared to just measuring in the linear regime (see Supplemental Material~VII for details). Note that the latter is an effect expected for increased dynamic-range measurements~\cite{Herbschleb2021}, it is not a direct effect of the low-frequency algorithm itself.

We exhibit the working of our algorithm at a field $2$ orders of magnitude lower than before, and in principle, it works at zero field. However, around zero field, it is in the nonlinear regime, and the sensitivity would decrease [contrary to a nonzero field offset, multiple readout phases are not possible in this case, see Fig.~\ref{Figure:Low field}(d,e)]. Measuring at a delay-dependent field offset, for a delay of $0.2$~ms it is a few tens of nT, a signal can be measured in the linear regime at a sensitivity about $1.5$ times worse compared to high-field measurements, as the states with a nitrogen nuclear spin of $0$ are not utilized (thus lowering the contrast) given the dependence on electrical field and strain. This is in principle sufficient for ultralow field NMR and for biomedical applications. Even at zero field, the field is only as zero as the signal that is measured. Nonetheless, from a theoretical point of view, the sensitivity at zero field while measuring a field with an amplitude of zero is negligible, since it is always at an extremum, which has zero gradient [Fig.~\ref{Figure:Low field}(a)]. However, a neat way to circumvent this is to apply circularly polarized MW pulses~\cite{Zheng2019}, also possible for large areas~\cite{Yaroshenko2020}, which allows to move the maximum gradient to zero field. Thus, with such technical additions, the algorithm itself works at the true zero field as well.

For demonstrating the measurement of synchronized signals, a NMR signal is chosen. Although here it is not our focus, we give a short comparison with previous N-V center NMR research~\cite{Glenn2018}. There, the high-frequency NMR spectrum for water is measured with an ensemble of N-V centers resulting in a line width of $9\pm1$~Hz with a signal frequency of approximately $3.8$~MHz for the free nuclear precession measurement. Opposed to the convention of standard NMR, they found their line width via a fit to the power spectrum, which decreases the line width to $\sqrt{\sqrt{2}-1}\approx64$\% of the conventional line width. The conventional line width for our measurement of a water sample is $1.6$~Hz (see Supplemental Material~VIII), hence in principle we show an improvement by an order of magnitude. However, note that their experiment is rather different, so this does not properly reflect the methods. Ultimately, for both methods, the line width is limited by the coherence time of the NMR sample in conjunction with the noise. Since at lower fields, narrower line widths are possible under similar conditions~\cite{McDermott2002}, there is significant potential for our low-frequency algorithm. Although the low-frequency NMR signal is created specifically to demonstrate the synchronized measurement, as potential application we consider low-field NMR with a N-V center near the surface~\cite{Watanabe2021} to sense chemicals or single molecules~\cite{Kost2015}.

For synchronized measurements, the sensitivity is frequency independent up to a maximum frequency. However, note that for the nonsynchronized measurements, even though the sensitivity is the same for every frequency as well, a single sequence is designed for a certain period only. By accumulating measurements over many periods, the sequence functions as a filter, where the shape depends on the number of accumulations and with maxima directly related to this period (for the base frequency and its harmonics up to a maximum). Thus, for the nonsynchronized case, the same sensitivity can be reached for any low frequency by changing the sequence accordingly.

Finally, we reflect on our method in a more general way. In principle, the measurement subsequence could be replaced by different options, and the analysis method could be replaced by alternatives, and low-frequency fields can be measured still. When it comes to the measurement subsequence, for example, pulsed ODMR could replace the Ramsey subsequence, while other promising sequences are less straightforward to use when repeating the full sequence over multiple periods, such as ones with MW frequency offsets~\cite{Vutha2015}. Many choices for the analysis method exist as well. For example, there are Fourier-based methods~\cite{Glenn2018} and harmonic inversion~\cite{Mandelshtam1997}, which require little prior information, where harmonic inversion is more susceptible to noise but has a great spectral resolution, while Fourier is rather intuitive. On the other hand, Bayesian inference~\cite{Hincks2018} requires decent prior knowledge, and excels in analyzing large data with underlying models. Fitting is somewhere in between: some prior knowledge is often helpful, it is resilient to noise, it has a decent spectral resolution, it is fairly intuitive, but the number of data points should be relatively small. The choice depends on the circumstances. In our case, we choose to use Ramsey, as it gives the best sensitivity, and we choose to analyze via fitting, which processes all data points at once, suitable for accumulated data, and in principle any periodic signal with known shape, such as a triangular wave, can be measured.

In conclusion, we demonstrate an algorithm with a frequency-independent sensitivity for measuring low-frequency fields. We show its working for ac magnetic fields, yielding a sensitivity of $9.4$~nT~Hz$^{-0.5}$ for a single N-V center for frequencies ranging from about $0.5$~kHz down to $1$~Hz, and it is expected that the sensitivity remains the same at even lower frequencies. The algorithm works for any periodic low-frequency field, as essentially it works as a quantum oscilloscope, and multiple parameters, such as phase and offset, can be determined. Moreover, the algorithm works at ultralow field, here demonstrated at $30$~nT, and it can be extended to zero field. As example of a synchronized signal, we measure low-frequency NMR spectra for deionized water and ethanol, showing line widths in the order of a Hz. This technique is promising for applications that require highly sensitive low-frequency quantum sensing with nanoscale resolution such as low-field NMR, magnetic resonance imaging, and diagnostic evaluation of integrated circuits.

\nocite{Maze2008,Jamonneau2016}

\begin{acknowledgments}
	The authors acknowledge the financial support from JST OPERA (No. JPMJOP1841), KAKENHI (No. 21H04653) and the Collaborative Research Program of ICR, Kyoto University (2021-114).
\end{acknowledgments}

\appendix*

\section{NMR signal generation}
The static magnetic field $B_0$ is generated by a thermally stabilized permanent magnet ($B_0=1.0$~T; Magriteck Spinsolve 43 Carbon). The rf coil for excitation and detection (swapped between via a Mini-Circuits ZYSWA-2-50DR+ switch) of nuclear spins is wound around the NMR glass tube and inserted to the NMR magnet ($3.1$-mm coil diameter; $8$-turn solenoid; $L=186$~nH). The sample volume surrounded by the coil is approximately $20$~\textmu l. The coil is tuned to the frequency of the proton spin ($44.145$~MHz) with a standard $LC$ circuit with variable tuning and matching capacitors and the quality factor of the coil is $210$.

The rf pulse is generated by an arbitrary waveform generator (Rigol DG 4102) and is typically approximately $1.5$~ms long with approximately $1$~mW power at the resonant frequency. The NMR signal from nuclear spins is first amplified by a factor of $100$ ($40$~dB) by a low-noise voltage amplifier (FEMTO GmbH DHPVA-201), and then down-converted to audiofrequencies by mixing with a reference rf signal with a double-balanced mixer (R\&K Co Ltd MX010-0S). This signal is further amplified by $20$~dB and filtered for frequencies below $1$~kHz with a second amplifier (Stanford Research SR560) before transferring to the coil around the diamond sample. This coil consists of three turns and has a conversion factor of $12.3$~\textmu T V$^{-1}$.

For the NMR samples, deionized water and ethanol are obtained from Fujifilm Wako Co and degassed prior to use.

%


\begin{thebibliography}{40}%
\makeatletter
\providecommand \@ifxundefined [1]{%
 \@ifx{#1\undefined}
}%
\providecommand \@ifnum [1]{%
 \ifnum #1\expandafter \@firstoftwo
 \else \expandafter \@secondoftwo
 \fi
}%
\providecommand \@ifx [1]{%
 \ifx #1\expandafter \@firstoftwo
 \else \expandafter \@secondoftwo
 \fi
}%
\providecommand \natexlab [1]{#1}%
\providecommand \enquote  [1]{``#1''}%
\providecommand \bibnamefont  [1]{#1}%
\providecommand \bibfnamefont [1]{#1}%
\providecommand \citenamefont [1]{#1}%
\providecommand \href@noop [0]{\@secondoftwo}%
\providecommand \href [0]{\begingroup \@sanitize@url \@href}%
\providecommand \@href[1]{\@@startlink{#1}\@@href}%
\providecommand \@@href[1]{\endgroup#1\@@endlink}%
\providecommand \@sanitize@url [0]{\catcode `\\12\catcode `\$12\catcode
  `\&12\catcode `\#12\catcode `\^12\catcode `\_12\catcode `\%12\relax}%
\providecommand \@@startlink[1]{}%
\providecommand \@@endlink[0]{}%
\providecommand \url  [0]{\begingroup\@sanitize@url \@url }%
\providecommand \@url [1]{\endgroup\@href {#1}{\urlprefix }}%
\providecommand \urlprefix  [0]{URL }%
\providecommand \Eprint [0]{\href }%
\providecommand \doibase [0]{https://doi.org/}%
\providecommand \selectlanguage [0]{\@gobble}%
\providecommand \bibinfo  [0]{\@secondoftwo}%
\providecommand \bibfield  [0]{\@secondoftwo}%
\providecommand \translation [1]{[#1]}%
\providecommand \BibitemOpen [0]{}%
\providecommand \bibitemStop [0]{}%
\providecommand \bibitemNoStop [0]{.\EOS\space}%
\providecommand \EOS [0]{\spacefactor3000\relax}%
\providecommand \BibitemShut  [1]{\csname bibitem#1\endcsname}%
\let\auto@bib@innerbib\@empty
\bibitem [{\citenamefont {Degen}\ \emph {et~al.}(2017)\citenamefont {Degen},
  \citenamefont {Reinhard},\ and\ \citenamefont {Cappellaro}}]{Degen2017}%
  \BibitemOpen
  \bibfield  {author} {\bibinfo {author} {\bibfnamefont {C.~L.}\ \bibnamefont
  {Degen}}, \bibinfo {author} {\bibfnamefont {F.}~\bibnamefont {Reinhard}},\
  and\ \bibinfo {author} {\bibfnamefont {P.}~\bibnamefont {Cappellaro}},\
  }\bibfield  {title} {\bibinfo {title} {Quantum sensing},\ }\href@noop {}
  {\bibfield  {journal} {\bibinfo  {journal} {Rev. Mod. Phys.}\ }\textbf
  {\bibinfo {volume} {89}},\ \bibinfo {pages} {035002} (\bibinfo {year}
  {2017})}\BibitemShut {NoStop}%
\bibitem [{\citenamefont {Rondin}\ \emph {et~al.}(2014)\citenamefont {Rondin},
  \citenamefont {Tetienne}, \citenamefont {Hingant}, \citenamefont {Roch},
  \citenamefont {Maletinsky},\ and\ \citenamefont {Jacques}}]{Rondin2014}%
  \BibitemOpen
  \bibfield  {author} {\bibinfo {author} {\bibfnamefont {L.}~\bibnamefont
  {Rondin}}, \bibinfo {author} {\bibfnamefont {J.-P.}\ \bibnamefont
  {Tetienne}}, \bibinfo {author} {\bibfnamefont {T.}~\bibnamefont {Hingant}},
  \bibinfo {author} {\bibfnamefont {J.-F.}\ \bibnamefont {Roch}}, \bibinfo
  {author} {\bibfnamefont {P.}~\bibnamefont {Maletinsky}},\ and\ \bibinfo
  {author} {\bibfnamefont {V.}~\bibnamefont {Jacques}},\ }\bibfield  {title}
  {\bibinfo {title} {Magnetometry with nitrogen-vacancy defects in diamond},\
  }\href@noop {} {\bibfield  {journal} {\bibinfo  {journal} {Rep. Prog. Phys.}\
  }\textbf {\bibinfo {volume} {77}},\ \bibinfo {pages} {056503} (\bibinfo
  {year} {2014})}\BibitemShut {NoStop}%
\bibitem [{\citenamefont {Herbschleb}\ \emph {et~al.}(2019)\citenamefont
  {Herbschleb}, \citenamefont {Kato}, \citenamefont {Maruyama}, \citenamefont
  {Danjo}, \citenamefont {Makino}, \citenamefont {Yamasaki}, \citenamefont
  {Ohki}, \citenamefont {Hayashi}, \citenamefont {Morishita}, \citenamefont
  {Fujiwara},\ and\ \citenamefont {Mizuochi}}]{Herbschleb2019}%
  \BibitemOpen
  \bibfield  {author} {\bibinfo {author} {\bibfnamefont {E.~D.}\ \bibnamefont
  {Herbschleb}}, \bibinfo {author} {\bibfnamefont {H.}~\bibnamefont {Kato}},
  \bibinfo {author} {\bibfnamefont {Y.}~\bibnamefont {Maruyama}}, \bibinfo
  {author} {\bibfnamefont {T.}~\bibnamefont {Danjo}}, \bibinfo {author}
  {\bibfnamefont {T.}~\bibnamefont {Makino}}, \bibinfo {author} {\bibfnamefont
  {S.}~\bibnamefont {Yamasaki}}, \bibinfo {author} {\bibfnamefont
  {I.}~\bibnamefont {Ohki}}, \bibinfo {author} {\bibfnamefont {K.}~\bibnamefont
  {Hayashi}}, \bibinfo {author} {\bibfnamefont {H.}~\bibnamefont {Morishita}},
  \bibinfo {author} {\bibfnamefont {M.}~\bibnamefont {Fujiwara}},\ and\
  \bibinfo {author} {\bibfnamefont {N.}~\bibnamefont {Mizuochi}},\ }\bibfield
  {title} {\bibinfo {title} {Ultra-long coherence times amongst
  room-temperature solid-state spins},\ }\href@noop {} {\bibfield  {journal}
  {\bibinfo  {journal} {Nat. Commun.}\ }\textbf {\bibinfo {volume} {10}},\
  \bibinfo {pages} {3766} (\bibinfo {year} {2019})}\BibitemShut {NoStop}%
\bibitem [{\citenamefont {Bar-Gill}\ \emph {et~al.}(2013)\citenamefont
  {Bar-Gill}, \citenamefont {Pham}, \citenamefont {Jarmola}, \citenamefont
  {Budker},\ and\ \citenamefont {Walsworth}}]{Bar-Gill2013}%
  \BibitemOpen
  \bibfield  {author} {\bibinfo {author} {\bibfnamefont {N.}~\bibnamefont
  {Bar-Gill}}, \bibinfo {author} {\bibfnamefont {L.~M.}\ \bibnamefont {Pham}},
  \bibinfo {author} {\bibfnamefont {A.}~\bibnamefont {Jarmola}}, \bibinfo
  {author} {\bibfnamefont {D.}~\bibnamefont {Budker}},\ and\ \bibinfo {author}
  {\bibfnamefont {R.~L.}\ \bibnamefont {Walsworth}},\ }\bibfield  {title}
  {\bibinfo {title} {Solid-state electronic spin coherence time approaching one
  second},\ }\href@noop {} {\bibfield  {journal} {\bibinfo  {journal} {Nat.
  Commun.}\ }\textbf {\bibinfo {volume} {4}},\ \bibinfo {pages} {1743}
  (\bibinfo {year} {2013})}\BibitemShut {NoStop}%
\bibitem [{\citenamefont {Stark}\ \emph {et~al.}(2017)\citenamefont {Stark},
  \citenamefont {Aharon}, \citenamefont {Unden}, \citenamefont {Louzon},
  \citenamefont {Huck}, \citenamefont {Retzker}, \citenamefont {Andersen},\
  and\ \citenamefont {Jelezko}}]{Stark2017}%
  \BibitemOpen
  \bibfield  {author} {\bibinfo {author} {\bibfnamefont {A.}~\bibnamefont
  {Stark}}, \bibinfo {author} {\bibfnamefont {N.}~\bibnamefont {Aharon}},
  \bibinfo {author} {\bibfnamefont {T.}~\bibnamefont {Unden}}, \bibinfo
  {author} {\bibfnamefont {D.}~\bibnamefont {Louzon}}, \bibinfo {author}
  {\bibfnamefont {A.}~\bibnamefont {Huck}}, \bibinfo {author} {\bibfnamefont
  {A.}~\bibnamefont {Retzker}}, \bibinfo {author} {\bibfnamefont {U.~L.}\
  \bibnamefont {Andersen}},\ and\ \bibinfo {author} {\bibfnamefont
  {F.}~\bibnamefont {Jelezko}},\ }\bibfield  {title} {\bibinfo {title}
  {Narrow-bandwidth sensing of high-frequency fields with continuous dynamical
  decoupling},\ }\href {https://doi.org/10.1038/s41467-017-01159-2} {\bibfield
  {journal} {\bibinfo  {journal} {Nat. Commun.}\ }\textbf {\bibinfo {volume}
  {8}},\ \bibinfo {pages} {1105} (\bibinfo {year} {2017})}\BibitemShut
  {NoStop}%
\bibitem [{\citenamefont {Meinel}\ \emph {et~al.}(2021)\citenamefont {Meinel},
  \citenamefont {Vorobyov}, \citenamefont {Yavkin}, \citenamefont {Dasari},
  \citenamefont {Sumiya}, \citenamefont {Onoda}, \citenamefont {Isoya},\ and\
  \citenamefont {Wrachtrup}}]{Meinel2021}%
  \BibitemOpen
  \bibfield  {author} {\bibinfo {author} {\bibfnamefont {J.}~\bibnamefont
  {Meinel}}, \bibinfo {author} {\bibfnamefont {V.}~\bibnamefont {Vorobyov}},
  \bibinfo {author} {\bibfnamefont {B.}~\bibnamefont {Yavkin}}, \bibinfo
  {author} {\bibfnamefont {D.}~\bibnamefont {Dasari}}, \bibinfo {author}
  {\bibfnamefont {H.}~\bibnamefont {Sumiya}}, \bibinfo {author} {\bibfnamefont
  {S.}~\bibnamefont {Onoda}}, \bibinfo {author} {\bibfnamefont
  {J.}~\bibnamefont {Isoya}},\ and\ \bibinfo {author} {\bibfnamefont
  {J.}~\bibnamefont {Wrachtrup}},\ }\bibfield  {title} {\bibinfo {title}
  {Heterodyne sensing of microwaves with a quantum sensor},\ }\href@noop {}
  {\bibfield  {journal} {\bibinfo  {journal} {Nat. Commun.}\ }\textbf {\bibinfo
  {volume} {12}},\ \bibinfo {pages} {2737} (\bibinfo {year}
  {2021})}\BibitemShut {NoStop}%
\bibitem [{\citenamefont {Acosta}\ \emph {et~al.}(2010)\citenamefont {Acosta},
  \citenamefont {Jarmola}, \citenamefont {Bauch},\ and\ \citenamefont
  {Budker}}]{Acosta2010a}%
  \BibitemOpen
  \bibfield  {author} {\bibinfo {author} {\bibfnamefont {V.~M.}\ \bibnamefont
  {Acosta}}, \bibinfo {author} {\bibfnamefont {A.}~\bibnamefont {Jarmola}},
  \bibinfo {author} {\bibfnamefont {E.}~\bibnamefont {Bauch}},\ and\ \bibinfo
  {author} {\bibfnamefont {D.}~\bibnamefont {Budker}},\ }\bibfield  {title}
  {\bibinfo {title} {Optical properties of the nitrogen-vacancy singlet levels
  in diamond},\ }\href@noop {} {\bibfield  {journal} {\bibinfo  {journal}
  {Phys. Rev. B}\ }\textbf {\bibinfo {volume} {82}},\ \bibinfo {pages} {201202}
  (\bibinfo {year} {2010})}\BibitemShut {NoStop}%
\bibitem [{\citenamefont {Schoenfeld}\ and\ \citenamefont
  {Harneit}(2011)}]{Schoenfeld2011}%
  \BibitemOpen
  \bibfield  {author} {\bibinfo {author} {\bibfnamefont {R.~S.}\ \bibnamefont
  {Schoenfeld}}\ and\ \bibinfo {author} {\bibfnamefont {W.}~\bibnamefont
  {Harneit}},\ }\bibfield  {title} {\bibinfo {title} {Real time magnetic field
  sensing and imaging using a single spin in diamond},\ }\href@noop {}
  {\bibfield  {journal} {\bibinfo  {journal} {Phys. Rev. Lett.}\ }\textbf
  {\bibinfo {volume} {106}},\ \bibinfo {pages} {030802} (\bibinfo {year}
  {2011})}\BibitemShut {NoStop}%
\bibitem [{\citenamefont {Clevenson}\ \emph {et~al.}(2015)\citenamefont
  {Clevenson}, \citenamefont {Trusheim}, \citenamefont {Teale}, \citenamefont
  {Schröder}, \citenamefont {Braje},\ and\ \citenamefont
  {Englund}}]{Clevenson2015}%
  \BibitemOpen
  \bibfield  {author} {\bibinfo {author} {\bibfnamefont {H.}~\bibnamefont
  {Clevenson}}, \bibinfo {author} {\bibfnamefont {M.~E.}\ \bibnamefont
  {Trusheim}}, \bibinfo {author} {\bibfnamefont {C.}~\bibnamefont {Teale}},
  \bibinfo {author} {\bibfnamefont {T.}~\bibnamefont {Schröder}}, \bibinfo
  {author} {\bibfnamefont {D.}~\bibnamefont {Braje}},\ and\ \bibinfo {author}
  {\bibfnamefont {D.}~\bibnamefont {Englund}},\ }\bibfield  {title} {\bibinfo
  {title} {Broadband magnetometry and temperature sensing with a light-trapping
  diamond waveguide},\ }\href@noop {} {\bibfield  {journal} {\bibinfo
  {journal} {Nat. Phys.}\ }\textbf {\bibinfo {volume} {11}},\ \bibinfo {pages}
  {393} (\bibinfo {year} {2015})}\BibitemShut {NoStop}%
\bibitem [{\citenamefont {Barry}\ \emph {et~al.}(2016)\citenamefont {Barry},
  \citenamefont {Turner}, \citenamefont {Schloss}, \citenamefont {Glenn},
  \citenamefont {Song}, \citenamefont {Lukin}, \citenamefont {Park},\ and\
  \citenamefont {Walsworth}}]{Barry2016}%
  \BibitemOpen
  \bibfield  {author} {\bibinfo {author} {\bibfnamefont {J.~F.}\ \bibnamefont
  {Barry}}, \bibinfo {author} {\bibfnamefont {M.~J.}\ \bibnamefont {Turner}},
  \bibinfo {author} {\bibfnamefont {J.~M.}\ \bibnamefont {Schloss}}, \bibinfo
  {author} {\bibfnamefont {D.~R.}\ \bibnamefont {Glenn}}, \bibinfo {author}
  {\bibfnamefont {Y.}~\bibnamefont {Song}}, \bibinfo {author} {\bibfnamefont
  {M.~D.}\ \bibnamefont {Lukin}}, \bibinfo {author} {\bibfnamefont
  {H.}~\bibnamefont {Park}},\ and\ \bibinfo {author} {\bibfnamefont {R.~L.}\
  \bibnamefont {Walsworth}},\ }\bibfield  {title} {\bibinfo {title} {Optical
  magnetic detection of single-neuron action potentials using quantum defects
  in diamond},\ }\href {https://doi.org/10.1073/pnas.1601513113} {\bibfield
  {journal} {\bibinfo  {journal} {Proc. Natl. Acad. Sci.}\ }\textbf {\bibinfo
  {volume} {113}},\ \bibinfo {pages} {14133} (\bibinfo {year}
  {2016})}\BibitemShut {NoStop}%
\bibitem [{\citenamefont {Schloss}\ \emph {et~al.}(2018)\citenamefont
  {Schloss}, \citenamefont {Barry}, \citenamefont {Turner},\ and\ \citenamefont
  {Walsworth}}]{Schloss2018}%
  \BibitemOpen
  \bibfield  {author} {\bibinfo {author} {\bibfnamefont {J.~M.}\ \bibnamefont
  {Schloss}}, \bibinfo {author} {\bibfnamefont {J.~F.}\ \bibnamefont {Barry}},
  \bibinfo {author} {\bibfnamefont {M.~J.}\ \bibnamefont {Turner}},\ and\
  \bibinfo {author} {\bibfnamefont {R.~L.}\ \bibnamefont {Walsworth}},\
  }\bibfield  {title} {\bibinfo {title} {Simultaneous broadband vector
  magnetometry using solid-state spins},\ }\href
  {https://doi.org/10.1103/PhysRevApplied.10.034044} {\bibfield  {journal}
  {\bibinfo  {journal} {Phys. Rev. Appl.}\ }\textbf {\bibinfo {volume} {10}},\
  \bibinfo {pages} {034044} (\bibinfo {year} {2018})}\BibitemShut {NoStop}%
\bibitem [{\citenamefont {Zhang}\ \emph {et~al.}(2021)\citenamefont {Zhang},
  \citenamefont {Shagieva}, \citenamefont {Widmann}, \citenamefont {K\"ubler},
  \citenamefont {Vorobyov}, \citenamefont {Kapitanova}, \citenamefont
  {Nenasheva}, \citenamefont {Corkill}, \citenamefont {Rhrle}, \citenamefont
  {Nakamura}, \citenamefont {Sumiya}, \citenamefont {Onoda}, \citenamefont
  {Isoya},\ and\ \citenamefont {Wrachtrup}}]{Zhang2021}%
  \BibitemOpen
  \bibfield  {author} {\bibinfo {author} {\bibfnamefont {C.}~\bibnamefont
  {Zhang}}, \bibinfo {author} {\bibfnamefont {F.}~\bibnamefont {Shagieva}},
  \bibinfo {author} {\bibfnamefont {M.}~\bibnamefont {Widmann}}, \bibinfo
  {author} {\bibfnamefont {M.}~\bibnamefont {K\"ubler}}, \bibinfo {author}
  {\bibfnamefont {V.}~\bibnamefont {Vorobyov}}, \bibinfo {author}
  {\bibfnamefont {P.}~\bibnamefont {Kapitanova}}, \bibinfo {author}
  {\bibfnamefont {E.}~\bibnamefont {Nenasheva}}, \bibinfo {author}
  {\bibfnamefont {R.}~\bibnamefont {Corkill}}, \bibinfo {author} {\bibfnamefont
  {O.}~\bibnamefont {Rhrle}}, \bibinfo {author} {\bibfnamefont
  {K.}~\bibnamefont {Nakamura}}, \bibinfo {author} {\bibfnamefont
  {H.}~\bibnamefont {Sumiya}}, \bibinfo {author} {\bibfnamefont
  {S.}~\bibnamefont {Onoda}}, \bibinfo {author} {\bibfnamefont
  {J.}~\bibnamefont {Isoya}},\ and\ \bibinfo {author} {\bibfnamefont
  {J.}~\bibnamefont {Wrachtrup}},\ }\bibfield  {title} {\bibinfo {title}
  {Diamond magnetometry and gradiometry towards subpicotesla dc field
  measurement},\ }\href@noop {} {\bibfield  {journal} {\bibinfo  {journal}
  {Phys. Rev. Appl.}\ }\textbf {\bibinfo {volume} {15}},\ \bibinfo {pages}
  {064075} (\bibinfo {year} {2021})}\BibitemShut {NoStop}%
\bibitem [{\citenamefont {Blanchard}\ \emph {et~al.}(2013)\citenamefont
  {Blanchard}, \citenamefont {Ledbetter}, \citenamefont {Theis}, \citenamefont
  {Butler}, \citenamefont {Budker},\ and\ \citenamefont
  {Pines}}]{Blanchard2013}%
  \BibitemOpen
  \bibfield  {author} {\bibinfo {author} {\bibfnamefont {J.~W.}\ \bibnamefont
  {Blanchard}}, \bibinfo {author} {\bibfnamefont {M.~P.}\ \bibnamefont
  {Ledbetter}}, \bibinfo {author} {\bibfnamefont {T.}~\bibnamefont {Theis}},
  \bibinfo {author} {\bibfnamefont {M.~C.}\ \bibnamefont {Butler}}, \bibinfo
  {author} {\bibfnamefont {D.}~\bibnamefont {Budker}},\ and\ \bibinfo {author}
  {\bibfnamefont {A.}~\bibnamefont {Pines}},\ }\bibfield  {title} {\bibinfo
  {title} {High-resolution zero-field {NMR} {J}-spectroscopy of aromatic
  compounds},\ }\href {https://doi.org/10.1021/ja312239v} {\bibfield  {journal}
  {\bibinfo  {journal} {J. Am. Chem. Soc.}\ }\textbf {\bibinfo {volume}
  {135}},\ \bibinfo {pages} {3607} (\bibinfo {year} {2013})}\BibitemShut
  {NoStop}%
\bibitem [{\citenamefont {Barskiy}\ \emph {et~al.}(2019)\citenamefont
  {Barskiy}, \citenamefont {Tayler}, \citenamefont {Marco-Rius}, \citenamefont
  {Kurhanewicz}, \citenamefont {Vigneron}, \citenamefont {Cikrikci},
  \citenamefont {Aydogdu}, \citenamefont {Reh}, \citenamefont {Pravdivtsev},
  \citenamefont {Hövener}, \citenamefont {Blanchard}, \citenamefont {Wu},
  \citenamefont {Budker},\ and\ \citenamefont {Pines}}]{Barskiy2019}%
  \BibitemOpen
  \bibfield  {author} {\bibinfo {author} {\bibfnamefont {D.~A.}\ \bibnamefont
  {Barskiy}}, \bibinfo {author} {\bibfnamefont {M.~C.~D.}\ \bibnamefont
  {Tayler}}, \bibinfo {author} {\bibfnamefont {I.}~\bibnamefont {Marco-Rius}},
  \bibinfo {author} {\bibfnamefont {J.}~\bibnamefont {Kurhanewicz}}, \bibinfo
  {author} {\bibfnamefont {D.~B.}\ \bibnamefont {Vigneron}}, \bibinfo {author}
  {\bibfnamefont {S.}~\bibnamefont {Cikrikci}}, \bibinfo {author}
  {\bibfnamefont {A.}~\bibnamefont {Aydogdu}}, \bibinfo {author} {\bibfnamefont
  {M.}~\bibnamefont {Reh}}, \bibinfo {author} {\bibfnamefont {A.~N.}\
  \bibnamefont {Pravdivtsev}}, \bibinfo {author} {\bibfnamefont {J.-B.}\
  \bibnamefont {Hövener}}, \bibinfo {author} {\bibfnamefont {J.~W.}\
  \bibnamefont {Blanchard}}, \bibinfo {author} {\bibfnamefont {T.}~\bibnamefont
  {Wu}}, \bibinfo {author} {\bibfnamefont {D.}~\bibnamefont {Budker}},\ and\
  \bibinfo {author} {\bibfnamefont {A.}~\bibnamefont {Pines}},\ }\bibfield
  {title} {\bibinfo {title} {Zero-field nuclear magnetic resonance of
  chemically exchanging systems},\ }\href
  {https://doi.org/10.1038/s41467-019-10787-9} {\bibfield  {journal} {\bibinfo
  {journal} {Nat. Commun.}\ }\textbf {\bibinfo {volume} {10}},\ \bibinfo
  {pages} {3002} (\bibinfo {year} {2019})}\BibitemShut {NoStop}%
\bibitem [{\citenamefont {Garcon}\ \emph {et~al.}(2019)\citenamefont {Garcon},
  \citenamefont {Blanchard}, \citenamefont {Centers}, \citenamefont {Figueroa},
  \citenamefont {Graham}, \citenamefont {Jackson~Kimball}, \citenamefont
  {Rajendran}, \citenamefont {Sushkov}, \citenamefont {Stadnik}, \citenamefont
  {Wickenbrock}, \citenamefont {Wu},\ and\ \citenamefont
  {Budker}}]{Garcon2019}%
  \BibitemOpen
  \bibfield  {author} {\bibinfo {author} {\bibfnamefont {A.}~\bibnamefont
  {Garcon}}, \bibinfo {author} {\bibfnamefont {J.~W.}\ \bibnamefont
  {Blanchard}}, \bibinfo {author} {\bibfnamefont {G.~P.}\ \bibnamefont
  {Centers}}, \bibinfo {author} {\bibfnamefont {N.~L.}\ \bibnamefont
  {Figueroa}}, \bibinfo {author} {\bibfnamefont {P.~W.}\ \bibnamefont
  {Graham}}, \bibinfo {author} {\bibfnamefont {D.~F.}\ \bibnamefont
  {Jackson~Kimball}}, \bibinfo {author} {\bibfnamefont {S.}~\bibnamefont
  {Rajendran}}, \bibinfo {author} {\bibfnamefont {A.~O.}\ \bibnamefont
  {Sushkov}}, \bibinfo {author} {\bibfnamefont {Y.~V.}\ \bibnamefont
  {Stadnik}}, \bibinfo {author} {\bibfnamefont {A.}~\bibnamefont
  {Wickenbrock}}, \bibinfo {author} {\bibfnamefont {T.}~\bibnamefont {Wu}},\
  and\ \bibinfo {author} {\bibfnamefont {D.}~\bibnamefont {Budker}},\
  }\bibfield  {title} {\bibinfo {title} {Constraints on bosonic dark matter
  from ultralow-field nuclear magnetic resonance},\ }\href
  {https://doi.org/10.1126/sciadv.aax4539} {\bibfield  {journal} {\bibinfo
  {journal} {Sci. Adv.}\ }\textbf {\bibinfo {volume} {5}},\ \bibinfo {pages}
  {eaax4539} (\bibinfo {year} {2019})}\BibitemShut {NoStop}%
\bibitem [{\citenamefont {Wu}\ \emph {et~al.}(2019)\citenamefont {Wu},
  \citenamefont {Blanchard}, \citenamefont {Centers}, \citenamefont {Figueroa},
  \citenamefont {Garcon}, \citenamefont {Graham}, \citenamefont {Kimball},
  \citenamefont {Rajendran}, \citenamefont {Stadnik}, \citenamefont {Sushkov},
  \citenamefont {Wickenbrock},\ and\ \citenamefont {Budker}}]{Wu2019}%
  \BibitemOpen
  \bibfield  {author} {\bibinfo {author} {\bibfnamefont {T.}~\bibnamefont
  {Wu}}, \bibinfo {author} {\bibfnamefont {J.~W.}\ \bibnamefont {Blanchard}},
  \bibinfo {author} {\bibfnamefont {G.~P.}\ \bibnamefont {Centers}}, \bibinfo
  {author} {\bibfnamefont {N.~L.}\ \bibnamefont {Figueroa}}, \bibinfo {author}
  {\bibfnamefont {A.}~\bibnamefont {Garcon}}, \bibinfo {author} {\bibfnamefont
  {P.~W.}\ \bibnamefont {Graham}}, \bibinfo {author} {\bibfnamefont {D.~F.~J.}\
  \bibnamefont {Kimball}}, \bibinfo {author} {\bibfnamefont {S.}~\bibnamefont
  {Rajendran}}, \bibinfo {author} {\bibfnamefont {Y.~V.}\ \bibnamefont
  {Stadnik}}, \bibinfo {author} {\bibfnamefont {A.~O.}\ \bibnamefont
  {Sushkov}}, \bibinfo {author} {\bibfnamefont {A.}~\bibnamefont
  {Wickenbrock}},\ and\ \bibinfo {author} {\bibfnamefont {D.}~\bibnamefont
  {Budker}},\ }\bibfield  {title} {\bibinfo {title} {Search for axionlike dark
  matter with a liquid-state nuclear spin comagnetometer},\ }\href
  {https://doi.org/10.1103/PhysRevLett.122.191302} {\bibfield  {journal}
  {\bibinfo  {journal} {Phys. Rev. Lett.}\ }\textbf {\bibinfo {volume} {122}},\
  \bibinfo {pages} {191302} (\bibinfo {year} {2019})}\BibitemShut {NoStop}%
\bibitem [{\citenamefont {McDermott}\ \emph {et~al.}(2002)\citenamefont
  {McDermott}, \citenamefont {Trabesinger}, \citenamefont {M{\"u}ck},
  \citenamefont {Hahn}, \citenamefont {Pines},\ and\ \citenamefont
  {Clarke}}]{McDermott2002}%
  \BibitemOpen
  \bibfield  {author} {\bibinfo {author} {\bibfnamefont {R.}~\bibnamefont
  {McDermott}}, \bibinfo {author} {\bibfnamefont {A.~H.}\ \bibnamefont
  {Trabesinger}}, \bibinfo {author} {\bibfnamefont {M.}~\bibnamefont
  {M{\"u}ck}}, \bibinfo {author} {\bibfnamefont {E.~L.}\ \bibnamefont {Hahn}},
  \bibinfo {author} {\bibfnamefont {A.}~\bibnamefont {Pines}},\ and\ \bibinfo
  {author} {\bibfnamefont {J.}~\bibnamefont {Clarke}},\ }\bibfield  {title}
  {\bibinfo {title} {Liquid-state {NMR} and scalar couplings in microtesla
  magnetic fields},\ }\href {https://doi.org/10.1126/science.1069280}
  {\bibfield  {journal} {\bibinfo  {journal} {Science}\ }\textbf {\bibinfo
  {volume} {295}},\ \bibinfo {pages} {2247} (\bibinfo {year}
  {2002})}\BibitemShut {NoStop}%
\bibitem [{\citenamefont {Hoult}\ and\ \citenamefont
  {Richards}(1975)}]{Hoult1975}%
  \BibitemOpen
  \bibfield  {author} {\bibinfo {author} {\bibfnamefont {D.~I.}\ \bibnamefont
  {Hoult}}\ and\ \bibinfo {author} {\bibfnamefont {R.~E.}\ \bibnamefont
  {Richards}},\ }\bibfield  {title} {\bibinfo {title} {Critical factors in the
  design of sensitive high resolution nuclear magnetic resonance
  spectrometers},\ }\href {https://doi.org/10.1098/rspa.1975.0104} {\bibfield
  {journal} {\bibinfo  {journal} {Proc. R. Soc. Lond. A}\ }\textbf {\bibinfo
  {volume} {344}},\ \bibinfo {pages} {311} (\bibinfo {year}
  {1975})}\BibitemShut {NoStop}%
\bibitem [{\citenamefont {Morris}(2017)}]{Morris2017}%
  \BibitemOpen
  \bibfield  {author} {\bibinfo {author} {\bibfnamefont {G.~A.}\ \bibnamefont
  {Morris}},\ }\bibinfo {title} {Encyclopedia of spectroscopy and spectrometry
  (third edition)}\ (\bibinfo  {publisher} {Academic Press},\ \bibinfo
  {address} {Oxford},\ \bibinfo {year} {2017})\ Chap.\ \bibinfo {chapter} {NMR
  Data Processing}\BibitemShut {NoStop}%
\bibitem [{\citenamefont {Dr\'eau}\ \emph {et~al.}(2011)\citenamefont
  {Dr\'eau}, \citenamefont {Lesik}, \citenamefont {Rondin}, \citenamefont
  {Spinicelli}, \citenamefont {Arcizet}, \citenamefont {Roch},\ and\
  \citenamefont {Jacques}}]{Dreau2011}%
  \BibitemOpen
  \bibfield  {author} {\bibinfo {author} {\bibfnamefont {A.}~\bibnamefont
  {Dr\'eau}}, \bibinfo {author} {\bibfnamefont {M.}~\bibnamefont {Lesik}},
  \bibinfo {author} {\bibfnamefont {L.}~\bibnamefont {Rondin}}, \bibinfo
  {author} {\bibfnamefont {P.}~\bibnamefont {Spinicelli}}, \bibinfo {author}
  {\bibfnamefont {O.}~\bibnamefont {Arcizet}}, \bibinfo {author} {\bibfnamefont
  {J.-F.}\ \bibnamefont {Roch}},\ and\ \bibinfo {author} {\bibfnamefont
  {V.}~\bibnamefont {Jacques}},\ }\bibfield  {title} {\bibinfo {title}
  {Avoiding power broadening in optically detected magnetic resonance of single
  {NV} defects for enhanced dc magnetic field sensitivity},\ }\href@noop {}
  {\bibfield  {journal} {\bibinfo  {journal} {Phys. Rev. B}\ }\textbf {\bibinfo
  {volume} {84}},\ \bibinfo {pages} {195204} (\bibinfo {year}
  {2011})}\BibitemShut {NoStop}%
\bibitem [{\citenamefont {Zheng}\ \emph {et~al.}(2019)\citenamefont {Zheng},
  \citenamefont {Xu}, \citenamefont {Iwata}, \citenamefont {Lenz},
  \citenamefont {Michl}, \citenamefont {Yavkin}, \citenamefont {Nakamura},
  \citenamefont {Sumiya}, \citenamefont {Ohshima}, \citenamefont {Isoya},
  \citenamefont {Wrachtrup}, \citenamefont {Wickenbrock},\ and\ \citenamefont
  {Budker}}]{Zheng2019}%
  \BibitemOpen
  \bibfield  {author} {\bibinfo {author} {\bibfnamefont {H.}~\bibnamefont
  {Zheng}}, \bibinfo {author} {\bibfnamefont {J.}~\bibnamefont {Xu}}, \bibinfo
  {author} {\bibfnamefont {G.~Z.}\ \bibnamefont {Iwata}}, \bibinfo {author}
  {\bibfnamefont {T.}~\bibnamefont {Lenz}}, \bibinfo {author} {\bibfnamefont
  {J.}~\bibnamefont {Michl}}, \bibinfo {author} {\bibfnamefont
  {B.}~\bibnamefont {Yavkin}}, \bibinfo {author} {\bibfnamefont
  {K.}~\bibnamefont {Nakamura}}, \bibinfo {author} {\bibfnamefont
  {H.}~\bibnamefont {Sumiya}}, \bibinfo {author} {\bibfnamefont
  {T.}~\bibnamefont {Ohshima}}, \bibinfo {author} {\bibfnamefont
  {J.}~\bibnamefont {Isoya}}, \bibinfo {author} {\bibfnamefont
  {J.}~\bibnamefont {Wrachtrup}}, \bibinfo {author} {\bibfnamefont
  {A.}~\bibnamefont {Wickenbrock}},\ and\ \bibinfo {author} {\bibfnamefont
  {D.}~\bibnamefont {Budker}},\ }\bibfield  {title} {\bibinfo {title}
  {Zero-field magnetometry based on nitrogen-vacancy ensembles in diamond},\
  }\href@noop {} {\bibfield  {journal} {\bibinfo  {journal} {Phys. Rev. Appl.}\
  }\textbf {\bibinfo {volume} {11}},\ \bibinfo {pages} {064068} (\bibinfo
  {year} {2019})}\BibitemShut {NoStop}%
\bibitem [{\citenamefont {Cerrillo}\ \emph {et~al.}(2021)\citenamefont
  {Cerrillo}, \citenamefont {Oviedo~Casado},\ and\ \citenamefont
  {Prior}}]{Cerrillo2021}%
  \BibitemOpen
  \bibfield  {author} {\bibinfo {author} {\bibfnamefont {J.}~\bibnamefont
  {Cerrillo}}, \bibinfo {author} {\bibfnamefont {S.}~\bibnamefont
  {Oviedo~Casado}},\ and\ \bibinfo {author} {\bibfnamefont {J.}~\bibnamefont
  {Prior}},\ }\bibfield  {title} {\bibinfo {title} {Low field nano-{NMR} via
  three-level system control},\ }\href@noop {} {\bibfield  {journal} {\bibinfo
  {journal} {Phys. Rev. Lett.}\ }\textbf {\bibinfo {volume} {126}},\ \bibinfo
  {pages} {220402} (\bibinfo {year} {2021})}\BibitemShut {NoStop}%
\bibitem [{\citenamefont {Ramsey}(1950)}]{Ramsey1950}%
  \BibitemOpen
  \bibfield  {author} {\bibinfo {author} {\bibfnamefont {N.~F.}\ \bibnamefont
  {Ramsey}},\ }\bibfield  {title} {\bibinfo {title} {A molecular beam resonance
  method with separated oscillating fields},\ }\href@noop {} {\bibfield
  {journal} {\bibinfo  {journal} {Phys. Rev.}\ }\textbf {\bibinfo {volume}
  {78}},\ \bibinfo {pages} {695} (\bibinfo {year} {1950})}\BibitemShut
  {NoStop}%
\bibitem [{\citenamefont {Gruber}\ \emph {et~al.}(1997)\citenamefont {Gruber},
  \citenamefont {Dr{\"a}benstedt}, \citenamefont {Tietz}, \citenamefont
  {Fleury}, \citenamefont {Wrachtrup},\ and\ \citenamefont
  {Borczyskowski}}]{Gruber1997}%
  \BibitemOpen
  \bibfield  {author} {\bibinfo {author} {\bibfnamefont {A.}~\bibnamefont
  {Gruber}}, \bibinfo {author} {\bibfnamefont {A.}~\bibnamefont
  {Dr{\"a}benstedt}}, \bibinfo {author} {\bibfnamefont {C.}~\bibnamefont
  {Tietz}}, \bibinfo {author} {\bibfnamefont {L.}~\bibnamefont {Fleury}},
  \bibinfo {author} {\bibfnamefont {J.}~\bibnamefont {Wrachtrup}},\ and\
  \bibinfo {author} {\bibfnamefont {C.~v.}\ \bibnamefont {Borczyskowski}},\
  }\bibfield  {title} {\bibinfo {title} {Scanning confocal optical microscopy
  and magnetic resonance on single defect centers},\ }\href
  {https://doi.org/10.1126/science.276.5321.2012} {\bibfield  {journal}
  {\bibinfo  {journal} {Science}\ }\textbf {\bibinfo {volume} {276}},\ \bibinfo
  {pages} {2012} (\bibinfo {year} {1997})}\BibitemShut {NoStop}%
\bibitem [{\citenamefont {Herbschleb}\ \emph {et~al.}(2021)\citenamefont
  {Herbschleb}, \citenamefont {Kato}, \citenamefont {Makino}, \citenamefont
  {Yamasaki},\ and\ \citenamefont {Mizuochi}}]{Herbschleb2021}%
  \BibitemOpen
  \bibfield  {author} {\bibinfo {author} {\bibfnamefont {E.~D.}\ \bibnamefont
  {Herbschleb}}, \bibinfo {author} {\bibfnamefont {H.}~\bibnamefont {Kato}},
  \bibinfo {author} {\bibfnamefont {T.}~\bibnamefont {Makino}}, \bibinfo
  {author} {\bibfnamefont {S.}~\bibnamefont {Yamasaki}},\ and\ \bibinfo
  {author} {\bibfnamefont {N.}~\bibnamefont {Mizuochi}},\ }\bibfield  {title}
  {\bibinfo {title} {Ultra-high dynamic range quantum measurement retaining its
  sensitivity},\ }\href {https://doi.org/10.1038/s41467-020-20561-x} {\bibfield
   {journal} {\bibinfo  {journal} {Nat. Commun.}\ }\textbf {\bibinfo {volume}
  {12}},\ \bibinfo {pages} {306} (\bibinfo {year} {2021})}\BibitemShut
  {NoStop}%
\bibitem [{\citenamefont {Taylor}\ \emph {et~al.}(2008)\citenamefont {Taylor},
  \citenamefont {Cappellaro}, \citenamefont {Childress}, \citenamefont {Jiang},
  \citenamefont {Budker}, \citenamefont {Hemmer}, \citenamefont {Yacoby},
  \citenamefont {Walsworth},\ and\ \citenamefont {Lukin}}]{Taylor2008}%
  \BibitemOpen
  \bibfield  {author} {\bibinfo {author} {\bibfnamefont {J.~M.}\ \bibnamefont
  {Taylor}}, \bibinfo {author} {\bibfnamefont {P.}~\bibnamefont {Cappellaro}},
  \bibinfo {author} {\bibfnamefont {L.}~\bibnamefont {Childress}}, \bibinfo
  {author} {\bibfnamefont {L.}~\bibnamefont {Jiang}}, \bibinfo {author}
  {\bibfnamefont {D.}~\bibnamefont {Budker}}, \bibinfo {author} {\bibfnamefont
  {P.~R.}\ \bibnamefont {Hemmer}}, \bibinfo {author} {\bibfnamefont
  {A.}~\bibnamefont {Yacoby}}, \bibinfo {author} {\bibfnamefont
  {R.}~\bibnamefont {Walsworth}},\ and\ \bibinfo {author} {\bibfnamefont
  {M.~D.}\ \bibnamefont {Lukin}},\ }\bibfield  {title} {\bibinfo {title}
  {High-sensitivity diamond magnetometer with nanoscale resolution},\
  }\href@noop {} {\bibfield  {journal} {\bibinfo  {journal} {Nat. Phys.}\
  }\textbf {\bibinfo {volume} {4}},\ \bibinfo {pages} {810} (\bibinfo {year}
  {2008})}\BibitemShut {NoStop}%
\bibitem [{Note1()}]{Note1}%
  \BibitemOpen
  \bibinfo {note} {See Supplemental Material for derivations and supportive
  information.}\BibitemShut {Stop}%
\bibitem [{\citenamefont {Kato}\ \emph {et~al.}(2016)\citenamefont {Kato},
  \citenamefont {Ogura}, \citenamefont {Makino}, \citenamefont {Takeuchi},\
  and\ \citenamefont {Yamasaki}}]{Kato2016}%
  \BibitemOpen
  \bibfield  {author} {\bibinfo {author} {\bibfnamefont {H.}~\bibnamefont
  {Kato}}, \bibinfo {author} {\bibfnamefont {M.}~\bibnamefont {Ogura}},
  \bibinfo {author} {\bibfnamefont {T.}~\bibnamefont {Makino}}, \bibinfo
  {author} {\bibfnamefont {D.}~\bibnamefont {Takeuchi}},\ and\ \bibinfo
  {author} {\bibfnamefont {S.}~\bibnamefont {Yamasaki}},\ }\bibfield  {title}
  {\bibinfo {title} {N-type control of single-crystal diamond films by
  ultra-lightly phosphorus doping},\ }\href@noop {} {\bibfield  {journal}
  {\bibinfo  {journal} {Appl. Phys. Lett.}\ }\textbf {\bibinfo {volume}
  {109}},\ \bibinfo {pages} {142102} (\bibinfo {year} {2016})}\BibitemShut
  {NoStop}%
\bibitem [{\citenamefont {Glenn}\ \emph {et~al.}(2018)\citenamefont {Glenn},
  \citenamefont {Bucher}, \citenamefont {Lee}, \citenamefont {Lukin},
  \citenamefont {Park},\ and\ \citenamefont {Walsworth}}]{Glenn2018}%
  \BibitemOpen
  \bibfield  {author} {\bibinfo {author} {\bibfnamefont {D.~R.}\ \bibnamefont
  {Glenn}}, \bibinfo {author} {\bibfnamefont {D.~B.}\ \bibnamefont {Bucher}},
  \bibinfo {author} {\bibfnamefont {J.}~\bibnamefont {Lee}}, \bibinfo {author}
  {\bibfnamefont {M.~D.}\ \bibnamefont {Lukin}}, \bibinfo {author}
  {\bibfnamefont {H.}~\bibnamefont {Park}},\ and\ \bibinfo {author}
  {\bibfnamefont {R.~L.}\ \bibnamefont {Walsworth}},\ }\bibfield  {title}
  {\bibinfo {title} {High-resolution magnetic resonance spectroscopy using a
  solid-state spin sensor},\ }\href@noop {} {\bibfield  {journal} {\bibinfo
  {journal} {Nature}\ }\textbf {\bibinfo {volume} {555}},\ \bibinfo {pages}
  {351} (\bibinfo {year} {2018})}\BibitemShut {NoStop}%
\bibitem [{\citenamefont {Hitchman}\ \emph {et~al.}(1998)\citenamefont
  {Hitchman}, \citenamefont {Lilley},\ and\ \citenamefont
  {Campbell}}]{Hitchman1998}%
  \BibitemOpen
  \bibfield  {author} {\bibinfo {author} {\bibfnamefont {A.~P.}\ \bibnamefont
  {Hitchman}}, \bibinfo {author} {\bibfnamefont {F.~E.~M.}\ \bibnamefont
  {Lilley}},\ and\ \bibinfo {author} {\bibfnamefont {W.~H.}\ \bibnamefont
  {Campbell}},\ }\bibfield  {title} {\bibinfo {title} {The quiet daily
  variation in the total magnetic field: global curves},\ }\href@noop {}
  {\bibfield  {journal} {\bibinfo  {journal} {Geophys. Res. Lett.}\ }\textbf
  {\bibinfo {volume} {25}},\ \bibinfo {pages} {2007} (\bibinfo {year}
  {1998})}\BibitemShut {NoStop}%
\bibitem [{\citenamefont {Bovey}(1967)}]{Bovey1967}%
  \BibitemOpen
  \bibfield  {author} {\bibinfo {author} {\bibfnamefont {F.~A.}\ \bibnamefont
  {Bovey}},\ }\href@noop {} {\emph {\bibinfo {title} {{NMR} data tables for
  organic compounds}}}\ (\bibinfo  {publisher} {Wiley Interscience},\ \bibinfo
  {address} {New York},\ \bibinfo {year} {1967})\BibitemShut {NoStop}%
\bibitem [{\citenamefont {Hahn}(1950)}]{Hahn1950a}%
  \BibitemOpen
  \bibfield  {author} {\bibinfo {author} {\bibfnamefont {E.~L.}\ \bibnamefont
  {Hahn}},\ }\bibfield  {title} {\bibinfo {title} {Spin echoes},\ }\href
  {https://doi.org/10.1103/PhysRev.80.580} {\bibfield  {journal} {\bibinfo
  {journal} {Phys. Rev.}\ }\textbf {\bibinfo {volume} {80}},\ \bibinfo {pages}
  {580} (\bibinfo {year} {1950})}\BibitemShut {NoStop}%
\bibitem [{\citenamefont {Mandelshtam}\ and\ \citenamefont
  {Taylor}(1997)}]{Mandelshtam1997}%
  \BibitemOpen
  \bibfield  {author} {\bibinfo {author} {\bibfnamefont {V.~A.}\ \bibnamefont
  {Mandelshtam}}\ and\ \bibinfo {author} {\bibfnamefont {H.~S.}\ \bibnamefont
  {Taylor}},\ }\bibfield  {title} {\bibinfo {title} {Harmonic inversion of time
  signals and its applications},\ }\href {https://doi.org/10.1063/1.475324}
  {\bibfield  {journal} {\bibinfo  {journal} {J. Chem. Phys.}\ }\textbf
  {\bibinfo {volume} {107}},\ \bibinfo {pages} {6756} (\bibinfo {year}
  {1997})}\BibitemShut {NoStop}%
\bibitem [{\citenamefont {Yaroshenko}\ \emph {et~al.}(2020)\citenamefont
  {Yaroshenko}, \citenamefont {Soshenko}, \citenamefont {Vorobyov},
  \citenamefont {Bolshedvorskii}, \citenamefont {Nenasheva}, \citenamefont
  {Kotel’nikov}, \citenamefont {Akimov},\ and\ \citenamefont
  {Kapitanova}}]{Yaroshenko2020}%
  \BibitemOpen
  \bibfield  {author} {\bibinfo {author} {\bibfnamefont {V.}~\bibnamefont
  {Yaroshenko}}, \bibinfo {author} {\bibfnamefont {V.}~\bibnamefont
  {Soshenko}}, \bibinfo {author} {\bibfnamefont {V.}~\bibnamefont {Vorobyov}},
  \bibinfo {author} {\bibfnamefont {S.}~\bibnamefont {Bolshedvorskii}},
  \bibinfo {author} {\bibfnamefont {E.}~\bibnamefont {Nenasheva}}, \bibinfo
  {author} {\bibfnamefont {I.}~\bibnamefont {Kotel’nikov}}, \bibinfo {author}
  {\bibfnamefont {A.}~\bibnamefont {Akimov}},\ and\ \bibinfo {author}
  {\bibfnamefont {P.}~\bibnamefont {Kapitanova}},\ }\bibfield  {title}
  {\bibinfo {title} {Circularly polarized microwave antenna for nitrogen
  vacancy centers in diamond},\ }\href {https://doi.org/10.1063/1.5129863}
  {\bibfield  {journal} {\bibinfo  {journal} {Rev. Sci. Instrum.}\ }\textbf
  {\bibinfo {volume} {91}},\ \bibinfo {pages} {035003} (\bibinfo {year}
  {2020})}\BibitemShut {NoStop}%
\bibitem [{\citenamefont {Watanabe}\ \emph {et~al.}(2021)\citenamefont
  {Watanabe}, \citenamefont {Nishikawa}, \citenamefont {Kato}, \citenamefont
  {Fujie}, \citenamefont {Fujiwara}, \citenamefont {Makino}, \citenamefont
  {Yamasaki}, \citenamefont {Herbschleb},\ and\ \citenamefont
  {Mizuochi}}]{Watanabe2021}%
  \BibitemOpen
  \bibfield  {author} {\bibinfo {author} {\bibfnamefont {A.}~\bibnamefont
  {Watanabe}}, \bibinfo {author} {\bibfnamefont {T.}~\bibnamefont {Nishikawa}},
  \bibinfo {author} {\bibfnamefont {H.}~\bibnamefont {Kato}}, \bibinfo {author}
  {\bibfnamefont {M.}~\bibnamefont {Fujie}}, \bibinfo {author} {\bibfnamefont
  {M.}~\bibnamefont {Fujiwara}}, \bibinfo {author} {\bibfnamefont
  {T.}~\bibnamefont {Makino}}, \bibinfo {author} {\bibfnamefont
  {S.}~\bibnamefont {Yamasaki}}, \bibinfo {author} {\bibfnamefont {E.~D.}\
  \bibnamefont {Herbschleb}},\ and\ \bibinfo {author} {\bibfnamefont
  {N.}~\bibnamefont {Mizuochi}},\ }\bibfield  {title} {\bibinfo {title}
  {Shallow {NV} centers augmented by exploiting n-type diamond},\ }\href
  {https://doi.org/https://doi.org/10.1016/j.carbon.2021.03.010} {\bibfield
  {journal} {\bibinfo  {journal} {Carbon}\ }\textbf {\bibinfo {volume} {178}},\
  \bibinfo {pages} {294} (\bibinfo {year} {2021})}\BibitemShut {NoStop}%
\bibitem [{\citenamefont {Kost}\ \emph {et~al.}(2015)\citenamefont {Kost},
  \citenamefont {Cai},\ and\ \citenamefont {Plenio}}]{Kost2015}%
  \BibitemOpen
  \bibfield  {author} {\bibinfo {author} {\bibfnamefont {M.}~\bibnamefont
  {Kost}}, \bibinfo {author} {\bibfnamefont {J.}~\bibnamefont {Cai}},\ and\
  \bibinfo {author} {\bibfnamefont {M.~B.}\ \bibnamefont {Plenio}},\ }\bibfield
   {title} {\bibinfo {title} {Resolving single molecule structures with
  {N}itrogen-vacancy centers in diamond},\ }\href@noop {} {\bibfield  {journal}
  {\bibinfo  {journal} {Sci. Rep.}\ }\textbf {\bibinfo {volume} {5}},\ \bibinfo
  {pages} {11007} (\bibinfo {year} {2015})}\BibitemShut {NoStop}%
\bibitem [{\citenamefont {Vutha}\ and\ \citenamefont
  {Hessels}(2015)}]{Vutha2015}%
  \BibitemOpen
  \bibfield  {author} {\bibinfo {author} {\bibfnamefont {A.~C.}\ \bibnamefont
  {Vutha}}\ and\ \bibinfo {author} {\bibfnamefont {E.~A.}\ \bibnamefont
  {Hessels}},\ }\bibfield  {title} {\bibinfo {title} {Frequency-offset
  separated oscillatory fields},\ }\href
  {https://doi.org/10.1103/PhysRevA.92.052504} {\bibfield  {journal} {\bibinfo
  {journal} {Phys. Rev. A}\ }\textbf {\bibinfo {volume} {92}},\ \bibinfo
  {pages} {052504} (\bibinfo {year} {2015})}\BibitemShut {NoStop}%
\bibitem [{\citenamefont {Hincks}\ \emph {et~al.}(2018)\citenamefont {Hincks},
  \citenamefont {Granade},\ and\ \citenamefont {Cory}}]{Hincks2018}%
  \BibitemOpen
  \bibfield  {author} {\bibinfo {author} {\bibfnamefont {I.}~\bibnamefont
  {Hincks}}, \bibinfo {author} {\bibfnamefont {C.}~\bibnamefont {Granade}},\
  and\ \bibinfo {author} {\bibfnamefont {D.~G.}\ \bibnamefont {Cory}},\
  }\bibfield  {title} {\bibinfo {title} {Statistical inference with quantum
  measurements: methodologies for nitrogen vacancy centers in diamond},\
  }\href@noop {} {\bibfield  {journal} {\bibinfo  {journal} {New J. Phys.}\
  }\textbf {\bibinfo {volume} {20}},\ \bibinfo {pages} {013022} (\bibinfo
  {year} {2018})}\BibitemShut {NoStop}%
\bibitem [{\citenamefont {Maze}\ \emph {et~al.}(2008)\citenamefont {Maze},
  \citenamefont {Stanwix}, \citenamefont {Hodges}, \citenamefont {Hong},
  \citenamefont {Taylor}, \citenamefont {Cappellaro}, \citenamefont {Jiang},
  \citenamefont {Dutt}, \citenamefont {Togan}, \citenamefont {Zibrov},
  \citenamefont {Yacoby}, \citenamefont {Walsworth},\ and\ \citenamefont
  {Lukin}}]{Maze2008}%
  \BibitemOpen
  \bibfield  {author} {\bibinfo {author} {\bibfnamefont {J.~R.}\ \bibnamefont
  {Maze}}, \bibinfo {author} {\bibfnamefont {P.~L.}\ \bibnamefont {Stanwix}},
  \bibinfo {author} {\bibfnamefont {J.~S.}\ \bibnamefont {Hodges}}, \bibinfo
  {author} {\bibfnamefont {S.}~\bibnamefont {Hong}}, \bibinfo {author}
  {\bibfnamefont {J.~M.}\ \bibnamefont {Taylor}}, \bibinfo {author}
  {\bibfnamefont {P.}~\bibnamefont {Cappellaro}}, \bibinfo {author}
  {\bibfnamefont {L.}~\bibnamefont {Jiang}}, \bibinfo {author} {\bibfnamefont
  {M.~V.~G.}\ \bibnamefont {Dutt}}, \bibinfo {author} {\bibfnamefont
  {E.}~\bibnamefont {Togan}}, \bibinfo {author} {\bibfnamefont {A.~S.}\
  \bibnamefont {Zibrov}}, \bibinfo {author} {\bibfnamefont {A.}~\bibnamefont
  {Yacoby}}, \bibinfo {author} {\bibfnamefont {R.~L.}\ \bibnamefont
  {Walsworth}},\ and\ \bibinfo {author} {\bibfnamefont {M.~D.}\ \bibnamefont
  {Lukin}},\ }\bibfield  {title} {\bibinfo {title} {Nanoscale magnetic sensing
  with an individual electronic spin in diamond},\ }\href
  {https://doi.org/10.1038/nature07279} {\bibfield  {journal} {\bibinfo
  {journal} {Nature}\ }\textbf {\bibinfo {volume} {455}},\ \bibinfo {pages}
  {644} (\bibinfo {year} {2008})}\BibitemShut {NoStop}%
\bibitem [{\citenamefont {Jamonneau}\ \emph {et~al.}(2016)\citenamefont
  {Jamonneau}, \citenamefont {Lesik}, \citenamefont {Tetienne}, \citenamefont
  {Alvizu}, \citenamefont {Mayer}, \citenamefont {Dr\'eau}, \citenamefont
  {Kosen}, \citenamefont {Roch}, \citenamefont {Pezzagna}, \citenamefont
  {Meijer}, \citenamefont {Teraji}, \citenamefont {Kubo}, \citenamefont
  {Bertet}, \citenamefont {Maze},\ and\ \citenamefont
  {Jacques}}]{Jamonneau2016}%
  \BibitemOpen
  \bibfield  {author} {\bibinfo {author} {\bibfnamefont {P.}~\bibnamefont
  {Jamonneau}}, \bibinfo {author} {\bibfnamefont {M.}~\bibnamefont {Lesik}},
  \bibinfo {author} {\bibfnamefont {J.~P.}\ \bibnamefont {Tetienne}}, \bibinfo
  {author} {\bibfnamefont {I.}~\bibnamefont {Alvizu}}, \bibinfo {author}
  {\bibfnamefont {L.}~\bibnamefont {Mayer}}, \bibinfo {author} {\bibfnamefont
  {A.}~\bibnamefont {Dr\'eau}}, \bibinfo {author} {\bibfnamefont
  {S.}~\bibnamefont {Kosen}}, \bibinfo {author} {\bibfnamefont {J.-F.}\
  \bibnamefont {Roch}}, \bibinfo {author} {\bibfnamefont {S.}~\bibnamefont
  {Pezzagna}}, \bibinfo {author} {\bibfnamefont {J.}~\bibnamefont {Meijer}},
  \bibinfo {author} {\bibfnamefont {T.}~\bibnamefont {Teraji}}, \bibinfo
  {author} {\bibfnamefont {Y.}~\bibnamefont {Kubo}}, \bibinfo {author}
  {\bibfnamefont {P.}~\bibnamefont {Bertet}}, \bibinfo {author} {\bibfnamefont
  {J.~R.}\ \bibnamefont {Maze}},\ and\ \bibinfo {author} {\bibfnamefont
  {V.}~\bibnamefont {Jacques}},\ }\bibfield  {title} {\bibinfo {title}
  {Competition between electric field and magnetic field noise in the
  decoherence of a single spin in diamond},\ }\href@noop {} {\bibfield
  {journal} {\bibinfo  {journal} {Phys. Rev. B}\ }\textbf {\bibinfo {volume}
  {93}},\ \bibinfo {pages} {024305} (\bibinfo {year} {2016})}\BibitemShut
  {NoStop}%
\end{thebibliography}
\end{document}


\title{Supplemental Material: Low-frequency quantum sensing}

\author{E. D. Herbschleb}
\affiliation{Institute for Chemical Research, Kyoto University, Gokasho, Uji-city, Kyoto 611-0011, Japan}
\author{I. Ohki}
\affiliation{Institute for Chemical Research, Kyoto University, Gokasho, Uji-city, Kyoto 611-0011, Japan}
\affiliation{Institute for Quantum Life Science, National Institutes for Quantum Science and Technology, Chiba 263-8555, Japan}
\author{K. Morita}
\affiliation{Institute for Chemical Research, Kyoto University, Gokasho, Uji-city, Kyoto 611-0011, Japan}
\author{Y. Yoshii}
\affiliation{Sumida Corporation, KDX Ginza East Building 7F, 3-7-2, Irifune, Chuo-ku, Tokyo, 104-0042, Japan}
\author{H. Kato}
\author{T. Makino}
\affiliation{National Institute of Advanced Industrial Science and Technology (AIST), Tsukuba, Ibaraki 305-8568, Japan}
\author{S. Yamasaki}
\affiliation{Kanazawa University, Kanazawa, Ishikawa 920-1192, Japan}
\author{N. Mizuochi}
\affiliation{Institute for Chemical Research, Kyoto University, Gokasho, Uji-city, Kyoto 611-0011, Japan}
\affiliation{Center for Spintronics Research Network, Kyoto University, Uji, Kyoto 611-0011, Japan}

\maketitle

\tableofcontents

\newpage

\section{Sensitivity in the linear regime, and the time delay for its optimum}
\label{Supp:optimum time delay}

Below, the sensitivity in the linear regime for the fitting-based algorithm is derived. As basic ac signal, a sinusoidal magnetic field is sensed. For clarity, only the sensitivities for the ac and dc fields are derived, but a similar approach can be taken for any fitting parameter (see Supplemental Material~\ref{Supp:sensitivity others}). In Supplemental Fig.~\ref{SFigure:algorithm defines}, some of the variables relevant for the derivation are illustrated.

\begin{figure}[h]
	\includegraphics{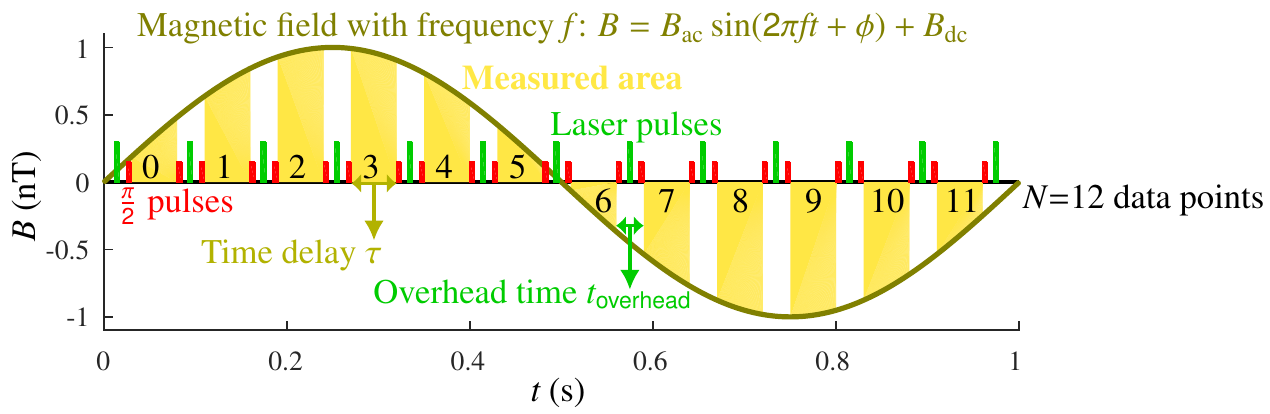}
	\caption{Illustration of various variables relevant to the algorithm.}
	\label{SFigure:algorithm defines}
\end{figure}

The sensitivity $\eta$ is defined as
\begin{equation}
\label{Equation:sensitivity}
\eta = dB_\textrm{min} \sqrt{T_\textrm{meas}},
\end{equation}
with $dB_\textrm{min}$ the smallest distinguishable field and $T_\textrm{meas}$ the measurement time. The former is generally defined as the uncertainty of the measured field $\sigma_B$~\cite{Maze2008,Herbschleb2019}, thus
\begin{equation}
\label{Equation:Bmin}
dB_\textrm{min} = \sigma_B.
\end{equation}

For our algorithm, these uncertainties are the standard errors of the corresponding fitting parameters. The readout signal $S$ at each data point (here the optical readout of the spin with the laser pulse) depends on the magnetic field $B$ (here represented as a single value instead of an integral as in the main text, thus for main-text frequency $\omega$ we have $\omega\tau\approx\omega_S$) as~\cite{Herbschleb2019}
\begin{equation}
\label{Equation:fitting function}
S = A\sin(\omega_S B+\theta)+O_S,
\end{equation}
with $A$ the amplitude, $\omega_S$ the frequency, $\theta$ the phase and $O_S$ the offset of the oscillation, caused by the rotation of the spin with increasing magnetic field amplitude. Here, the time $t$ dependence of this magnetic field $B$ is
\begin{equation}
\label{Equation:ac field}
B = B_\textrm{ac}\sin(2\pi f t+\phi)+B_\textrm{dc},
\end{equation}
with $B_\textrm{ac}$ the field amplitude, $f$ the frequency, $\phi$ the phase and $B_\textrm{dc}$ the constant field offset. Thus, the total fitting function is
\begin{equation}
\label{Equation:fitting function complete}
S = A\sin(\omega_S \left[B_\textrm{ac}\sin(2\pi f t+\phi)+B_\textrm{dc}\right] +\theta)+O_S.
\end{equation}
The uncertainties required for computing the sensitivities are the standard errors of the $B_\textrm{ac}$ fitting coefficient for ac, and of the $B_\textrm{dc}$ fitting coefficient for dc.

The standard errors for the fitting coefficients, $\sigma_\textrm{coef}$, follow from
\begin{equation}
\label{Equation:fitting coefficient standard error}
\sigma_\textrm{coef} = R \sqrt{\frac{\textrm{diag}\left(C\right)}{D}},
\end{equation}
with $R$ the residual norm, $C$ the covariance matrix and $D$ the degree of freedom. These are
\begin{equation}
\label{Equation:RCD}
\begin{aligned}
R &= \sqrt{\sum_{i=0}^{N-1}\left(F_i-f_i\right)^2} \\
C &= \left(J'\times J\right)^{-1} \\
D &= N-\#\textrm{coef},
\end{aligned}
\end{equation}
with $N$ the number of data points, $F_i$ the fitted result and $f_i$ the corresponding data point, $J$ the Jacobian for the fitting function, and $\#\textrm{coef}$ the number of fitting coefficients.

Lacking theoretical measurements, the residual norm can be estimated with the standard deviation $\sigma_1$ of a single data point as
\begin{equation}
\label{Equation:R}
R=\sqrt{\sum_{i=0}^{N-1}\left(F_i-f_i\right)^2}\approx \sqrt{N-1}\sigma_1.
\end{equation}
For a single N-V center, this uncertainty is shot-noise limited and given by~\cite{Herbschleb2019}
\begin{equation}
\label{Equation:shotnoise}
\sigma_1\approx\sigma_\textrm{SN}=\frac{1}{\sqrt{MN_\textrm{ph}}},
\end{equation}
with $M$ the number of iterations of the sequence, and $N_\textrm{ph}$ the average number of photons counted per readout pulse.

The Jacobian is
\begin{equation}
\label{Equation:Jacobian}
J=\left[{\begin{array}{ccc}
	\frac{\partial F}{\partial c_0}\left(t_0\right) & \cdots & \frac{\partial F}{\partial c_{\#\textrm{coef}-1}}\left(t_0\right) \\
	\vdots & \ddots & \vdots \\
	\frac{\partial F}{\partial c_0}\left(t_{N-1}\right) & \cdots & \frac{\partial F}{\partial c_{\#\textrm{coef}-1}}\left(t_{N-1}\right) \\
	\end{array}}\right],
\end{equation}
where $c_j$ are the fitting coefficients and $t_i$ the times when the data points are measured.

In the linear regime (for the measured signal $S$), $\sin(x)\approx x$, thus the fitting functions simplifies to
\begin{equation}
\label{Equation:fitting function linear regime}
S \approx A\left(\omega_S \left[B_\textrm{ac}\sin(2\pi f t+\phi)+B_\textrm{dc}\right] +\theta\right)+O_S = A\omega_S B_\textrm{ac}\sin(2\pi f t+\phi)+A\omega_S B_\textrm{dc} + A\theta+O_S.
\end{equation}
The latter shows why the coefficients of Supplemental Eq.~(\ref{Equation:fitting function}) are calibration constants: otherwise the field parameters cannot be determined uniquely. These can be computed from the N-V center's parameters (such as $T_2^*$) and the fixed time delay $\tau$ chosen for the measurement~\cite{Herbschleb2019}.

Only fitting for the desired ac and dc components of the field (so $\#\textrm{coef}=2$ with $B_\textrm{ac}$ the first coefficient $c_0$ and $B_\textrm{dc}$ the second coefficient $c_1$), the Jacobian gives
\begin{equation}
\label{Equation:Jacobian linear}
J=\left[{\begin{array}{cc}
	A\omega_S\sin(2\pi f t_0+\phi) & A\omega_S \\
	\vdots & \vdots \\
	A\omega_S\sin(2\pi f t_{N-1}+\phi) & A\omega_S \\
	\end{array}}\right],
\end{equation}
and the matrix product with its transpose
\begin{equation}
\label{Equation:JacJac linear}
J'\times J=A^2\omega_S^2\left[{\begin{array}{cc}
	\sum_{i=0}^{N-1} \sin[2](2\pi f t_i+\phi) & \sum_{i=0}^{N-1} \sin(2\pi f t_i+\phi) \\
	\sum_{i=0}^{N-1} \sin(2\pi f t_i+\phi) & N \\
	\end{array}}\right].
\end{equation}
The sums can be approximated with the knowledge that the measurement is performed during a single period. Thus the sines sum to zero, while for the squared sine the sum gives
\begin{equation}
\label{Equation:JacJac approximation linear}
\sum_{i=0}^{N-1} \sin[2](2\pi f t_i+\phi) \approx fN\int_{0}^{\frac{1}{f}}\sin[2](2\pi f t+\phi)dt = \frac{fN}{2}\int_{0}^{\frac{1}{f}}1-\cos(4\pi f t+2\phi)dt=\frac{N}{2},
\end{equation}
since the cosine integrates to zero over multiple whole periods. Remember that the time step $\Delta t=1/\left(fN\right)$. Now, the matrix product can be approximated as
\begin{equation}
\label{Equation:JacJac linear period}
J'\times J\approx A^2\omega_S^2\left[{\begin{array}{cc}
	\frac{N}{2} & 0 \\
	0 & N \\
	\end{array}}\right],
\end{equation}
and thus the covariance matrix is approximated by
\begin{equation}
\label{Equation:covariance linear period}
C=\left(J'\times J\right)^{-1}\approx\frac{1}{A^2\omega_S^2} \left[{\begin{array}{cc}
		\frac{2}{N} & 0 \\
		0 & \frac{1}{N} \\
\end{array}}\right].
\end{equation}

To simplify the following equations, two variables are described with others. The measurement time is the number of iterations $M$ times the sequence time $t_\textrm{seq}$ which is a single period of the signal which has frequency $f$, thus
\begin{equation}
\label{Equation:measurement time}
T_\textrm{meas}=M t_\textrm{seq}=\frac{M}{f},
\end{equation}
and the number of data points $N$ follows from the time delay $\tau$ between the $\pi/2$ pulses as
\begin{equation}
\label{Equation:data points}
N=\frac{1}{f\left(\tau+t_\textrm{overhead}\right)},
\end{equation}
with $t_\textrm{overhead}$ the overhead time in the sequence (laser pulses, MW pulses, waiting times).

Combining all above, the sensitivity for dc (the second fitting coefficient) is
\begin{equation}
\label{Equation:sensitivity dc}
\eta_\textrm{dc}=\sigma_\textrm{dc}\sqrt{T_\textrm{meas}}=\sqrt{N-1}\frac{1}{\sqrt{MN_\textrm{ph}}}\sqrt{\frac{\frac{1}{A^2\omega_S^2}\frac{1}{N}}{N-2}}\sqrt{\frac{M}{f}} \approx \frac{\sqrt{\tau+t_\textrm{overhead}}}{A\omega_S\sqrt{N_\textrm{ph}}},
\end{equation}
where the latter approximation holds for $N-1\approx N-2$, thus for large $N$. This formula happens to be exactly the same as the one for the sensitivity derived for the standard dc measurement in~\cite{Herbschleb2019}.

The sensitivity for ac (the first fitting coefficient) has just two differences with the dc one. Firstly, its value on the diagonal of the covariance matrix is two times larger. Secondly, its $\omega_S$ is slightly different (for details of its computation, see Supplemental Note~5 of~\cite{Herbschleb2019})
\begin{equation}
\label{Equation:frequency in field}
\omega_{S\textrm{, ac}}= \frac{g\mu_\textrm{B}}{\hbar}\int_{\frac{1}{4f}-\frac{\tau}{2}}^{\frac{1}{4f}+\frac{\tau}{2}}\sin(2\pi ft)dt \approx \frac{g\mu_\textrm{B}}{\hbar}\tau=\omega_{S\textrm{, dc}},
\end{equation}
with $g$ the g-factor of the electron spin, $\mu_\textrm{B}$ the Bohr magneton, and $\hbar$ reduced Planck's constant. The approximation becomes better for smaller $\tau$ with respect to $1/f$: it is rather accurate for low frequencies. Thus, the sensitivity for ac is
\begin{equation}
\label{Equation:sensitivity ac}
\eta_\textrm{ac}\approx\sqrt{2}\eta_\textrm{dc}.
\end{equation}

\begin{figure}[htbp]
	\includegraphics{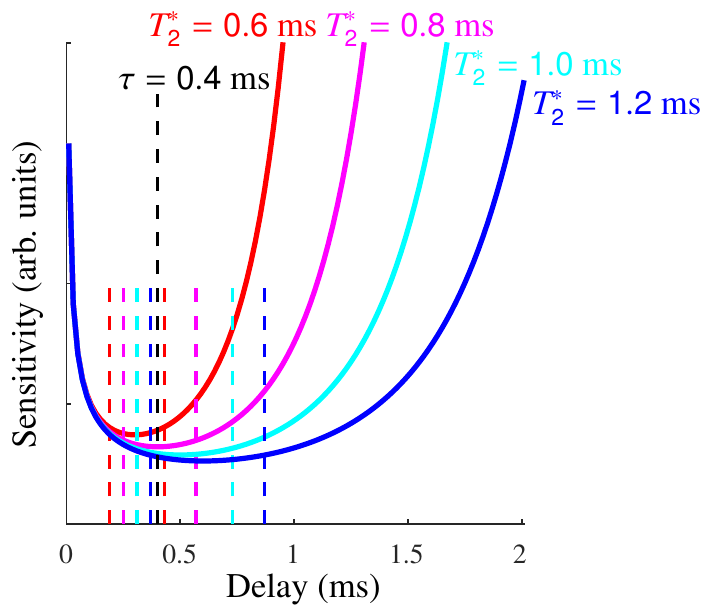}
	\caption{Calculated sensitivity vs delay for several $T_2^*$ (red $0.6$ ms, magenta $0.8$ ms, cyan $1.0$ ms and blue $1.2$ ms). The vertical color-coded dashed lines indicate the values for the delay for which the sensitivity remains within $10\%$ of the best value. The vertical black dashed line indicates the delay chosen in the main text ($\tau=0.4$ ms): it is within this $10\%$ for a range of apparent $T_2^*$s (see Supplemental Material~\ref{Supp:T2S} for details).}
	\label{SFigure:sensitivity vs delay}
\end{figure}

Finally, the optimum time delay $\tau_\textrm{optimum}$ is straightforward to find, since the formulae are equivalent shape-wise to the ones given in Supplementary Note~5 of~\cite{Herbschleb2019}. Thus, the optimum, which was derived there, is the same, which is repeated here for reference:
\begin{equation}
\label{Equation:optimum delay}
\tau_\textrm{optimum} =
\begin{cases}
T_2^*\sqrt[n]{\frac{1}{2n}} &\text{for}\ t_\textrm{overhead}\ll \tau_\textrm{optimum} \\
T_2^*\sqrt[n]{\frac{1}{n}} &\text{for}\ t_\textrm{overhead}\gg \tau_\textrm{optimum}.
\end{cases}
\end{equation}
Please note that $T_2^*$ is the inhomogeneous dephasing time and $n$ the power of its exponential. For single N-V centers, as used throughout this paper, $t_\textrm{overhead}\ll \tau_\textrm{optimum}$. Supplemental Fig.~\ref{SFigure:sensitivity vs delay} plots the calculated sensitivity for this case for several $T_2^*$s to justify the choice for the fixed delay in the main text of $0.4$ ms for the sensitivity measurements.

\newpage

\section{Notes on $T_2^*$}
\label{Supp:T2S}

As explained in Supplementary Note~2 of~\cite{Herbschleb2019}, the apparent (measured) inhomogeneous dephasing time $T_2^*$ is lower than the actual $T_2^*$ (limited by the sample) due to environmental effects. In an ideal system, these would be absent, and the apparent $T_2^*$ and the actual $T_2^*$ would be equal. However, since the experimental setup is hardly ideal, these external effects do lower the apparent $T_2^*$ with measurement time. This is visualized in Supplemental Fig.~\ref{SFigure:T2S}(a): for longer measurement times, the apparent $T_2^*$ decreases, while a short-time measurement with a gap~\cite{Herbschleb2019} indicates that the actual $T_2^*$ is about $1.21_{-0.08}^{+0.09}$ ms [Supplemental Fig.~\ref{SFigure:T2S}(b)].

Ideally, the optimum time delay (see Supplemental Material~\ref{Supp:optimum time delay}) should be about $T_2^*/2$. Since measuring the sensitivities of low-frequency signals requires long measurement times, it is chosen to use a time delay somewhat below the optimum, since, as shown in Supplemental Fig.~\ref{SFigure:sensitivity vs delay}, this only slightly affects the sensitivity for apparent $T_2^*$s within a reasonable range for the measured N-V center.

\begin{figure}[htbp]
	\includegraphics{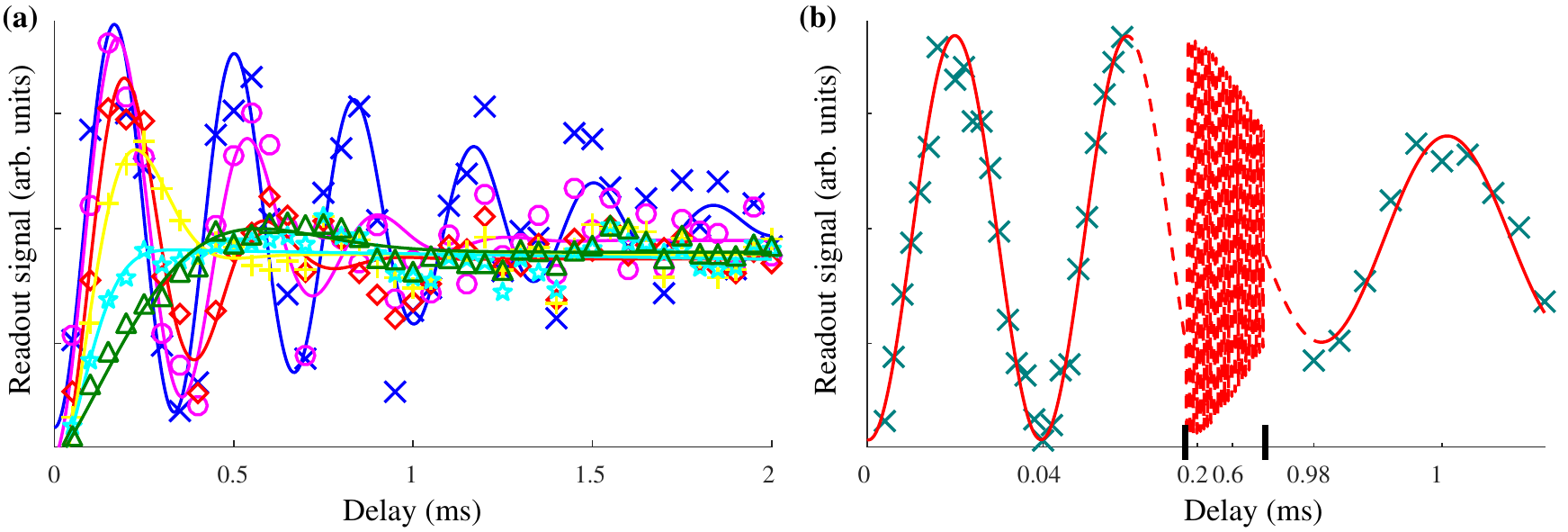}
	\caption{$T_2^*$ measurements. (a) $T_2^*$ measurement results taken at a relatively unstable time for increasing measurement time (blue crosses $\sim$$7$ minutes with $T_{2\textrm{, fit}}^*\approx 1.2$ ms, magenta circles $\sim$$18$ minutes with $T_{2\textrm{, fit}}^*\approx 0.62$ ms, red diamonds $\sim$$32$ minutes with $T_{2\textrm{, fit}}^*\approx 0.45$ ms, yellow pluses $\sim$$1$ hour with $T_{2\textrm{, fit}}^*\approx 0.26$ ms, cyan pentagons $\sim$$2$ hours with $T_{2\textrm{, fit}}^*\approx 0.16$ ms, dark green triangles $\sim$$4$ hours with $T_{2\textrm{, fit}}^*\approx 0.47$ ms). Please note the gradual change in frequency and the non-flat behavior of the long-time measurements, indicating a beat probably exists (caused by the slow change in frequency over time during the measurement). This affects the fitting for the long-time measurements as well. The data from main text Fig.~2(a) are taken at a stable time, so even though the measurement time is about an hour, the apparent $T_2^*$ is about $1$ ms. (b) Gapped $T_2^*$ measurement (explained in~\cite{Herbschleb2019}) lasting $10$ minutes giving $T_2^*=1.21_{-0.09}^{+0.10}$ ms; please note the breaks on the horizontal axis.}
	\label{SFigure:T2S}
\end{figure}

\newpage

\section{Sensitivity vs frequency}
\label{Supp:frequency dependence}

Using the formulae from Supplemental Material~\ref{Supp:optimum time delay}, but without the approximation of Supplemental Eq.~(\ref{Equation:frequency in field}), the sensitivity is optimized numerically for each frequency of the signal by choosing the number of data points $N$ that gives the best sensitivity. Initially, there is no restriction on the delay, the minimum number of data points is four [for four variables of Supplemental Eq.~(\ref{Equation:ac field})] and overhead time is neglected. The result is shown in Supplemental Fig.~\ref{SFigure:sensitivity vs frequency}(a) with blue and dark blue lines. The difference between the sensitivity for dc and low-frequency ac is $\sqrt{2}$, as expected from Supplemental Eq.~(\ref{Equation:sensitivity ac}). Moreover, the optimum delay at low frequencies is consistent with Supplemental Eq.~(\ref{Equation:optimum delay}).

The result for taking into account the overhead time (laser pulse, waiting times, MW pulses) that is required for the measurements in the main text is drawn with magenta and purple lines. As expected, the sensitivity is slightly worse, and more significantly worse for higher frequencies. The latter is because the delay becomes shorter given that at least $4$ data points are required, meaning the overhead time becomes a more significant part of the sequence.

For main text Fig.~3, the maximum delay is restricted to $0.4$ ms, while the minimum number of data points is still $4$, and of course the overhead time is taken into account. This worsens the sensitivity, as visualized by the red and dark red lines in Supplemental Fig.~\ref{SFigure:sensitivity vs frequency}(a), but it remains within $10\%$ of the optimum as designed (see Supplemental Fig.~\ref{SFigure:sensitivity vs delay}).

\begin{figure}[htbp]
	\includegraphics{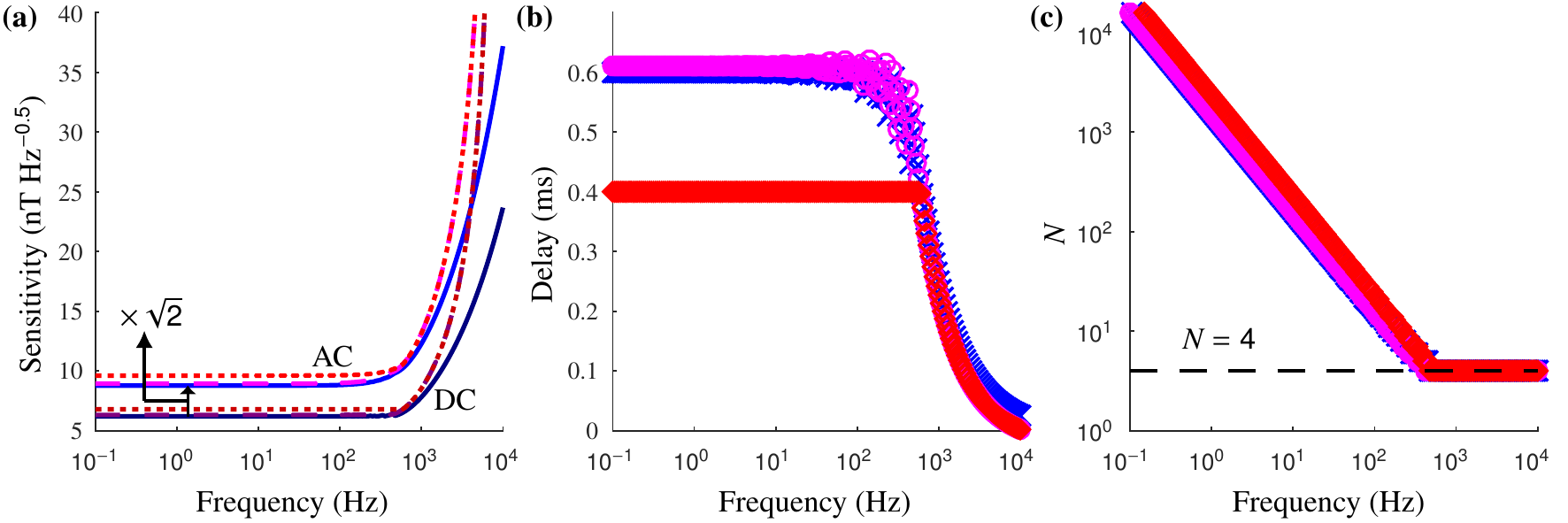}
	\caption{Sensitivity vs frequency for $T_2^*=1.2$~ms. (a) Lower darker lines indicate the dc sensitivity, while upper brighter lines give the ac sensitivity for each frequency. Blue and dark blue lines are for numerically optimized delays without taking overhead into account, magenta and purple dashed lines are for numerically optimized delays while taking overhead into account, and red and dark red dotted lines are for numerically optimized delays with a maximum of $0.4$~ms while taking overhead into account, as used in the main text for the sensitivity measurements. (b) Time delay between the $\pi/2$ pulses in the sequence for the same three scenarios as in (a) plotted with blue crosses, magenta circles and red diamonds respectively. As visible, the latter is restricted to a maximum delay of $0.4$ ms. (c) Number of data points $N$ during each period (see Supplemental Fig.~\ref{SFigure:algorithm defines}) for the same three scenarios as in (a) plotted with blue crosses, magenta circles and red diamonds respectively.}
	\label{SFigure:sensitivity vs frequency}
\end{figure}

\newpage

\section{Sensitivity of other fitting parameters}
\label{Supp:sensitivity others}

\begin{figure}[htbp]
	\includegraphics{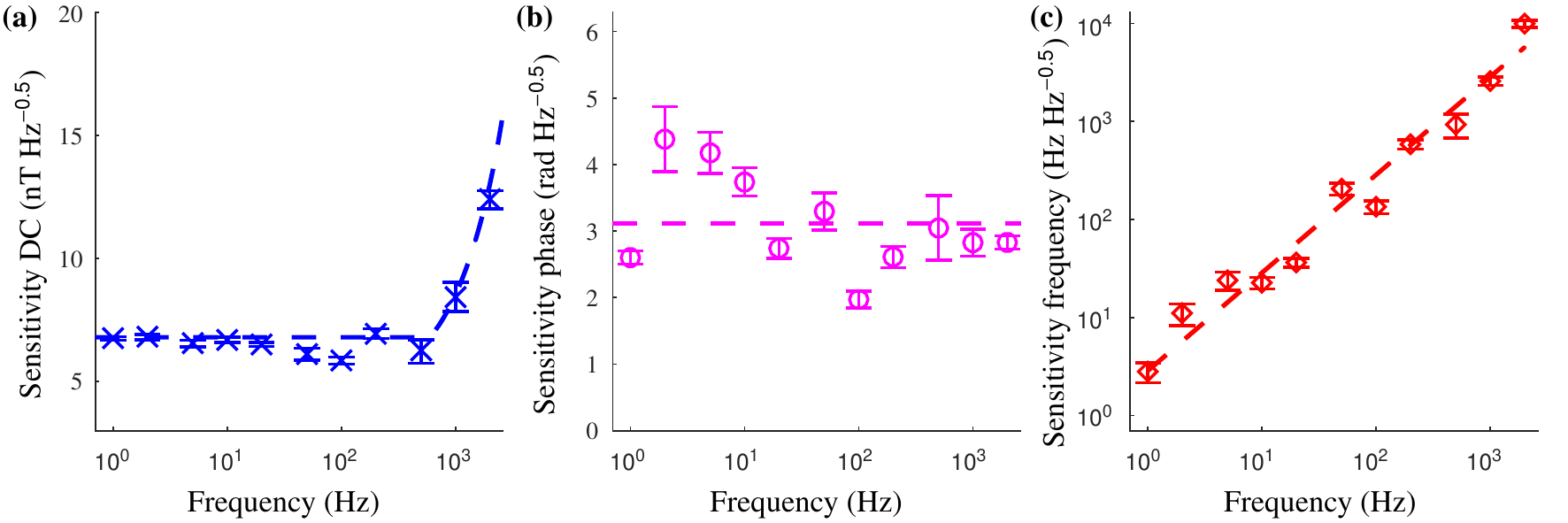}
	\caption{Sensitivities of other fitting parameters. (a) Sensitivity $\eta_\textrm{dc}$ for the dc component: data points with blue crosses (error bars indicate standard errors), calculated result [see Supplemental Fig.~\ref{SFigure:sensitivity vs frequency}(a)] with blue dashed line. (b) Sensitivity $\eta_\phi$ for the phase: data points with magenta circles (error bars indicate standard errors), the horizontal dashed line is a guide to the eye. Please refer to the text below for the explanation. (c) Sensitivity $\eta_f$ for the frequency: data points with red diamonds (error bars indicate standard errors), the diagonal dashed line is a fit to $\eta_f=Cf^{1}$ with $C$ a constant. Please refer to the text below for the explanation.}
	\label{SFigure:sensitivities of rest}
\end{figure}

For completeness, since four variables are fitted to the data (all of Supplemental Eq.~(\ref{Equation:ac field}), thus $B_\textrm{ac}$, $f$, $\phi$ and $B_\textrm{dc}$), but only the ac component is explored in the main text, Supplemental Fig.~\ref{SFigure:sensitivities of rest} displays the sensitivities for dc field $B_\textrm{dc}$, for phase $\phi$ and for frequency $f$. Please note that the latter is only functional when a measurement with synchronization is used. The dc sensitivity follows the expected behavior from calculation (see Supplemental Materials~\ref{Supp:optimum time delay}~and~\ref{Supp:frequency dependence}).

The phase sensitivity $\eta_\phi$ seems to be fairly frequency independent. However, please note that this independence, and the sensitivity looking as being $\pi$, are coincidences. The Jacobian for the phase coefficient $\phi$ [see Supplemental Eqs.~(\ref{Equation:Jacobian})~and~(\ref{Equation:fitting function linear regime})] only is
\begin{equation}
\label{Equation:Jacobian phase}
J=\left[{\begin{array}{c}
		A\omega_S B_\textrm{ac}\cos(2\pi f t_0+\phi)\\
		\vdots \\
		A\omega_S B_\textrm{ac}\cos(2\pi f t_{N-1}+\phi)\\
\end{array}}\right].
\end{equation}
This is rather similar to the Jacobian for the ac coefficient, and following equivalent steps one finds
\begin{equation}
\label{Equation:sensitivity phase}
\eta_\phi\approx\frac{\eta_\textrm{ac}}{B_\textrm{ac}}.
\end{equation}
Since $\eta_\textrm{ac}\approx9$~nT~Hz$^{-0.5}$ and the applied field $B_\textrm{ac}\approx3$~nT for the sensitivity measurements, $\eta_\phi\approx3$~rad~Hz$^{-0.5}$, as Supplemental Fig.~\ref{SFigure:sensitivities of rest}(b) confirms. Given the lower sensitivity for the ac field at higher frequencies, a two times higher field was applied for $f=2$~kHz, which thus scales the sensitivity for the phase, making it look like a straight line.

The frequency dependence of the frequency sensitivity $\eta_f$ is slightly more complicated. The Jacobian for the frequency coefficient $f$ [see Supplemental Eqs.~(\ref{Equation:Jacobian})~and~(\ref{Equation:fitting function linear regime})] only is
\begin{equation}
\label{Equation:Jacobian frequency}
J=\left[{\begin{array}{c}
		2\pi t_0 A\omega_S B_\textrm{ac}\cos(2\pi f t_0+\phi)\\
		\vdots \\
		2\pi t_{N-1} A\omega_S B_\textrm{ac}\cos(2\pi f t_{N-1}+\phi)\\
\end{array}}\right].
\end{equation}
Thus, the matrix product with its transpose is
\begin{equation}
	\label{Equation:JacJac linear frequency}
	J'\times J=\left(2\pi\right)^2 A^2 \omega_S^2 B_\textrm{ac}^2 \sum_{i=0}^{N-1} t_i^2 \cos[2](2\pi f t_i+\phi)=\left(2\pi\right)^2 A^2 \omega_S^2 B_\textrm{ac}^2 \textrm{SUM}.
\end{equation}
Approximating $\textrm{SUM}$ over a single period gives
\begin{equation}
\label{Equation:JacJac approximation linear frequency}
	\textrm{SUM}=\sum_{i=0}^{N-1} t_i^2 \cos[2](2\pi f t_i+\phi) \approx fN\int_{0}^{\frac{1}{f}}t^2\cos[2](2\pi f t+\phi)dt = \frac{fN}{2}\int_{0}^{\frac{1}{f}}t^2\left(1+\cos(4\pi f t+2\phi)\right)dt.
\end{equation}
Integrating by parts twice results in
\begin{equation}
\label{Equation:JacJac approximation linear frequency result}
\begin{aligned}
	\textrm{SUM} & = \left[\frac{fN}{4}\left\{\frac{t^3}{3}+\frac{1}{4\pi f}\left(t^2\sin(4\pi f t+2\phi)+\frac{2}{4\pi f}\left[t\cos(4\pi f t+2\phi)-\frac{1}{4\pi f}\sin(4\pi f t+2\phi)\right]\right)\right\}\right]_0^{\frac{1}{f}} \\ & =
	\frac{N}{4 f^2}\left\{\frac{1}{3}+\frac{1}{4\pi}\sin(2\phi)+\frac{2}{\left(4\pi\right)^2}\cos(2\phi)\right\}.
\end{aligned}
\end{equation}
Applying Supplemental Eq.~(\ref{Equation:fitting coefficient standard error}), with $N$ as given in Supplemental Eq.~(\ref{Equation:data points}), the uncertainty $\sigma_f\propto f^{3/2}/B_\textrm{ac}$. Since the sensitivity multiplies the uncertainty by $\sqrt{T_\textrm{meas}}=\sqrt{M/f}$ [see Supplemental Eq.~(\ref{Equation:measurement time})], $\eta_f\propto f/B_\textrm{ac}$, which is reflected in Supplemental Fig.~\ref{SFigure:sensitivities of rest}(c). Please note the following points.

Firstly, just like for the phase, the last point's measured sensitivity is relatively better compared to the other points given the higher field applied when measuring at this frequency. However, here, to focus on the frequency dependency, it is scaled accordingly in the graph.

Secondly, the uncertainty is calculated for measuring a single period. When measuring multiple periods $p$, the uncertainty scales as $p^{-3/2}$ and the sensitivity as $p^{-1}$. Since in a fixed time, the number of periods is inversely proportional to the frequency, when measuring multiple frequencies within this chosen time length, the sensitivity for each frequency is roughly the same.

Finally for completeness, although we won't go into detail here, this frequency sensitivity is purely based on fitting. On the other hand, when solely looking at the frequency, meaning if the signal has a known amplitude or phase (one suffices, but the known-phase one is more sensitive by $\pi$ and has a linear response), the accumulation increases the information about the signal. The reason is that a signal with a frequency for which the measurement is designed has the expected amplitude and phase, while even small changes in the frequency alter the amplitude and phase; the more accumulations, the larger the change. For a non-decaying periodic signal, the sensitivity of such a measurement is the same as for a non-accumulated measurement. (Both sensitivities are inversely proportional to the amplitude of the field still, just like in the fitting way described above.) The downside is the (much) smaller range of frequencies detectable, depending on the number of accumulations. Please note that this is merely true for the non-synchronized case.
\newpage

\section{Low field}
\label{Supp:Low field}

Canceling the field in the $z$-direction for the N-V center is performed in three steps. At zero field, the energy levels of the $\ket{-1}$ and $\ket{+1}$ spin states of the N-V center align; only with a non-zero field they move apart due to Zeeman splitting. Here, the nitrogen atom is the $^{14}$N isotope, so there are two times three peaks (hyperfine splitting) that need to overlap (if no other factors would play a role); the spin states are written as $\ket{\textrm{N-V electron spin, N nuclear spin}}$. Firstly, pulsed ODMR measurements are performed to find the characteristic $2$:$1$:$2$ valley depths, caused by the near overlap of the $\ket{-1,-1}$ and $\ket{+1,+1}$ energy levels at the lower frequency and the $\ket{-1,+1}$ and $\ket{+1,-1}$ energy levels at the higher frequency [Supplemental Fig.~\ref{SFigure:low field}(a,b)]. These have double the contrast since, as opposed to the center $\ket{-1,0}$ and $\ket{+1,0}$ energy level transitions, there are two different states of the nitrogen nuclear spin that allow a transition. The higher the contrast is at the side peaks, the closer we are to zero field. The accuracy depends on the power broadening of the microwave field, so a rough estimate is a fraction of the line width of the peaks (since when the difference is the exact line width, the valley depth would not have doubled). Therefore, after this step, we are in the order of \textmu Ts away from zero field.

Next, we measure various FIDs for the valley at the lower frequency with a small detuning. As long as the energy levels do not overlap exactly, there should be two frequencies visible in the FID signal, representing the two nearby energy levels. At first, while initially also increasing the time span of the FID sequence, the field is reduced towards zero by aiming for a single peak in the Fourier transform, indicating a single frequency remains [Supplemental Fig.~\ref{SFigure:low field}(c,f) and (d,g)]. At this point, the field is roughly within the Fourier resolution (here $1.7$~kHz) away from zero field (so $\sim$$60/2=30$~nT, taking into account that both energy levels change). When these frequencies are roughly just within the Fourier resolution, the apparent $T_2^*$ is rather short due to the effect of beating~\cite{Herbschleb2019}. (Actually, it depends on the coherence time, but since the FID length is chosen depending on the coherence time, looking at the Fourier resolution reaches the same conclusion.) By maximizing the apparent $T_2^*$, it is possible to get closer to zero field [Supplemental Fig.~\ref{SFigure:low field}(e,h)]. The cyan shaded region in main text Fig.~4(a) indicates the range of fields that includes the zero field based on these FID measurements. Please note that for the final FID fine-tuning, a small bias voltage is applied to the coil around the N-V center, thus this region is not around $0$~nT, which is when no voltage is applied.

Finally, a calibration measurement indicates the location of zero field with the highest accuracy. The reason is that, irrespective of the detuning during this measurement, the zero field will be at an extremum [Supplemental Fig.~\ref{SFigure:low field}(i-k)]. As long as a single extremum exists within the field range that includes the zero field (given the previous FID measurements), the zero field is found. The precision depends on the uncertainty of the phase of the fitting function, and for the measurement plotted in Fig.~4(a) of the main text, it is $0.7$~nT (longer measurements decrease this uncertainty). It is important to aim for zero detuning based on the final FID measurement, as only a precise detuning of zero gives the maximum contrast: for a small non-zero detuning the contrast lowers [Supplemental Fig.~\ref{SFigure:low field}(k)]. The higher the contrast is, the larger the maximum gradient is, and thus the better the sensitivity of the sensor will be.

\begin{figure}[htbp]
	\includegraphics{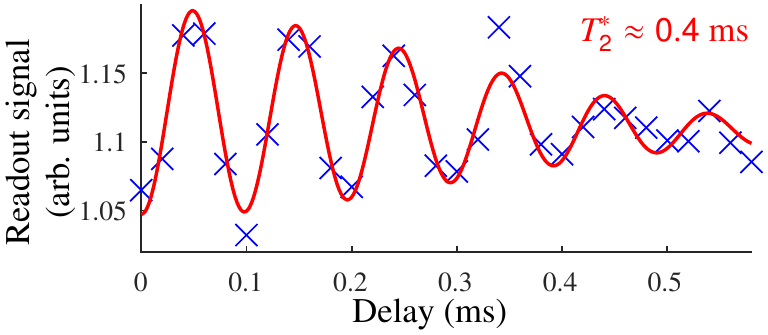}
	\caption{FID measurement at a background field of $\sim$$1.1$~mT.}
	\label{SFigure:high field}
\end{figure}

\begin{figure}[!htbp]
	\includegraphics{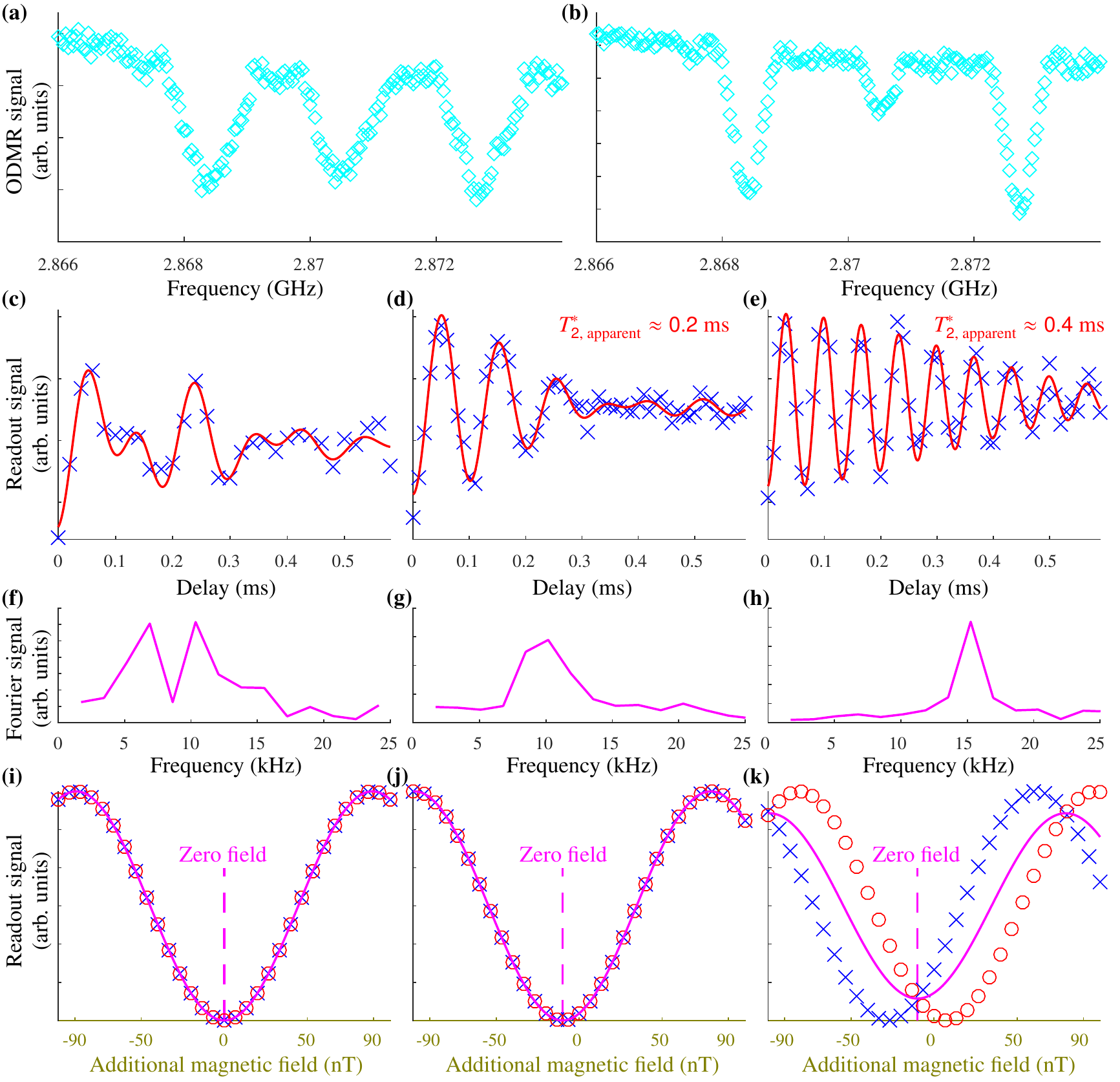}
	\caption{Zero field locating. (a) Pulsed ODMR spectrum close to zero field. (b) Pulsed ODMR spectrum sufficiently close to zero field to reduce the nuclear spin dependency at the outer frequencies, thus doubling their contrast. The center valley is slightly split due to electric field or strain, thus lowering its contrast. (c-e) FID measurements for background fields increasingly closer to zero field. For reference, the $T_2^*$s in (d) and (e) are found while fitting a single frequency. (f-h) Fourier spectra for (c-e). (i-k) Simulated calibration measurements (blue crosses and red circles for each of the two nearby transitions, magenta line for the resulting combination) for (i) a perfectly canceled field and no detuning, for (j) a remaining background field of $-10$~nT and no detuning, and for (k) a remaining background field of $-10$~nT and a detuning of $500$~Hz. Please note that for non-zero detuning, the contrast decreases (it will oscillate).}
	\label{SFigure:low field}
\end{figure}

When preparing the measurements at low field, we considered that the reason for the relatively low $T_2^*$ of the N-V center measured at first for this experiment is that at low fields, it was found that the coherence time decreased~\cite{Jamonneau2016}. Hence, we performed the measurements with this N-V center, considering this was a downside of measuring at low field. As the zero field does not remain for long after calibration, the sensitivity measurements had to be as short as possible, so a shorter delay would have been chosen either way, meaning that the lower $T_2^*$ did not affect the sensitivity in the measurement [main text Fig.~4(c)] that much (but it would when maximizing the sensitivity of course). Afterwards (as canceling the field is a time-consuming practice), we verified the $T_2^*$ at higher field ($\sim$$1.1$ mT, Supplemental Fig.~\ref{SFigure:high field}). It turns out that the $T_2^*$ of the N-V center does not increase at higher magnetic field, thus the relatively low $T_2^*$ appears to be just due to an unlucky choice. When getting closer to zero field, the apparent $T_2^*$ lowers [for example in Supplemental Fig.~\ref{SFigure:low field}(d)], while after reaching the Fourier resolution, the apparent $T_2^*$ increases again until it is at its actual $T_2^*$ at zero field [close to Supplemental Fig.~\ref{SFigure:low field}(e)].

\clearpage

\section{Alternative method}
\label{Supp:alternative method}

The Fourier transforms of Supplemental Material~\ref{Supp:LW} suggest an alternative way to find the amplitude of the ac signal by looking at the magnitude of the Fourier transformed signal. The transform for an example sinusoidal field with a frequency of $5$~Hz and an amplitude of $3.1$~nT (Supplemental Fig.~\ref{SFigure:Fourier}(a), this amplitude is used in many of the experiments of the main text) is plotted in Supplemental Fig.~\ref{SFigure:Fourier}(b). As a single period is measured generally, the frequency does not matter, as the resulting spectrum looks the same (a single peak around the first non-zero frequency). The transform is scaled with the number of data points, such that the peak magnitude gives the amplitude of the field. Fitting the Lorentzian from Supplemental Eq.~(\ref{Equation:Lorentzian}), this time without a constraint on the maximum value, gives for the amplitude $B_\textrm{ac}=3.1_{-0.1}^{+0.2}$~nT.

An example for a measured field with the same frequency and amplitude is given in Supplemental Fig.~\ref{SFigure:Fourier}(c). The time-domain fit (the algorithm from the main text) gives for the amplitude $B_\textrm{ac}=3.1_{-0.2}^{+0.2}$~nT. The Fourier transform is given in Supplemental Fig.~\ref{SFigure:Fourier}(d), which fit gives $B_\textrm{ac}=2.3_{-0.8}^{+24.1}$~nT.

There are several points to note here. First, since there is only a single point in the peak, finding the magnitude via a fit is challenging, as the error bounds indicate. Alternatively, one could just take the maximum data point itself as the amplitude, but this would not work for frequencies that are in between the data points, as is possible with synchronized measurements such as in Supplemental Fig.~\ref{SFigure:line widths}. As implied in Supplemental Material~\ref{Supp:LW}, the amplitude in freely fitting its data was rather inaccurate. Second, although the actual value is rather off ($2.3$~nT instead of $3.1$~nT), this is likely related to the shape of the field. The proposed fitting method automatically solves this by computing the integrals while fitting, thus some kind of correction is required for the Fourier amplitude, which would likely depend on the shape of the field (so it would be frequency dependent). When working with windowed Fourier transforms, this probably requires an additional correction as well.

Overall, since each data point in the time-domain reflects the amplitude of the field and allows for straightforwardly taking the shape into account, fitting in the time-domain accurately gives the amplitude of the low-frequency field. Although generally only a few points are part of the peak in the frequency domain, perhaps a similar performance is attainable still, but it seems less straightforward.

\begin{figure}[htbp]
	\includegraphics{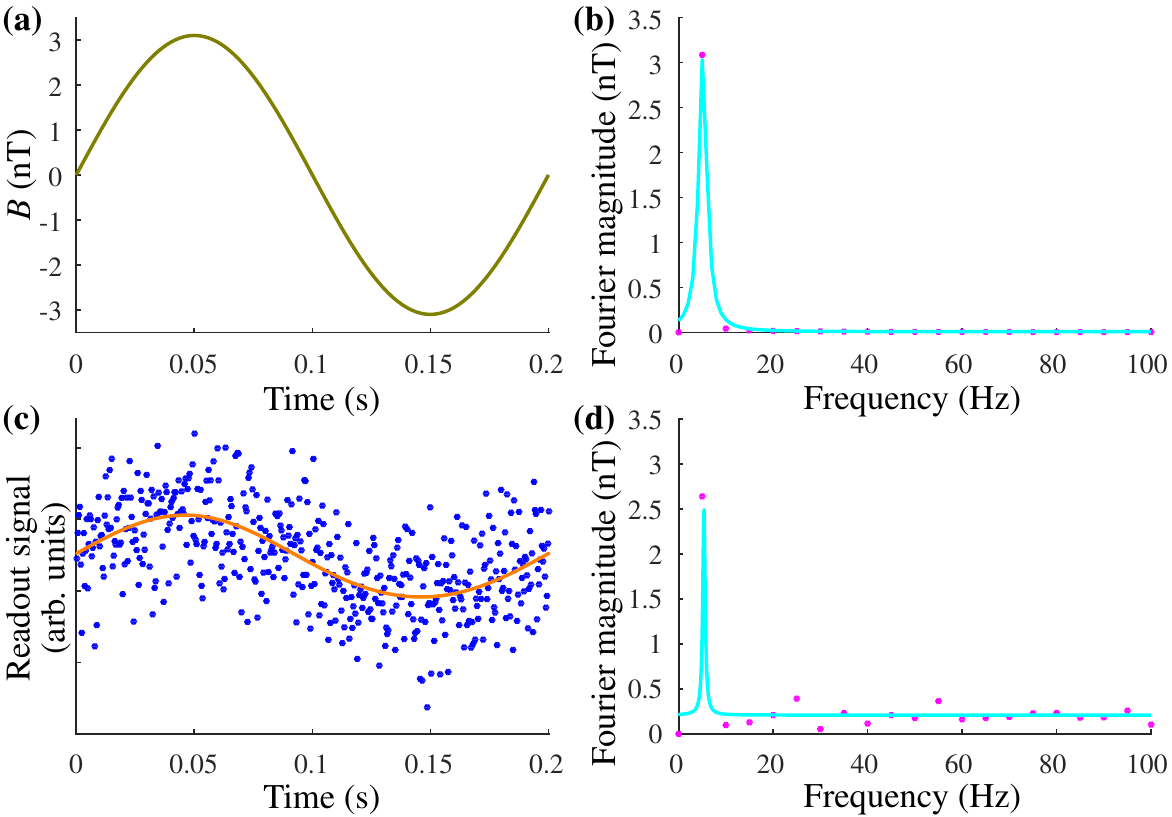}
	\caption{Finding the amplitude of a field in the time and frequency domains. (a) Example field with an amplitude of $3.1$~nT and a frequency of $5$~Hz. (b) Magnitude of the Fourier transform of (a), scaled with the number of data points so that the magnitude reflects the amplitude of the field (magenta points, fit with cyan line); only the low-frequency part is shown. The Lorentzian fit gives $B_\textrm{ac}=3.1_{-0.1}^{+0.2}$~nT. (c) Measurement data (blue dots) in the time-domain, which fit (orange line) gives for the amplitude $B_\textrm{ac}=3.1_{-0.2}^{+0.2}$~nT. (d) As (b), but for the field converted from the data points of (c). The fit gives $B_\textrm{ac}=2.3_{-0.8}^{+24.1}$~nT.}
	\label{SFigure:Fourier}
\end{figure}

\newpage

\section{Sensitivity in the non-linear regime: high dynamic-range}
\label{Supp:DR}

Utilizing the non-linear regime increases the dynamic range of the measurement. In this regime, as opposed to Supplemental Material~\ref{Supp:optimum time delay}, the approximation of Supplemental Eq.~(\ref{Equation:fitting function linear regime}) is not valid. Thus, the Jacobian becomes
\begin{equation}
\label{Equation:Jacobian non-linear}
J=\left[{\begin{array}{cc}
	A\omega_S\sin(2\pi f t_0+\phi)\cos(\omega_S \left[B_\textrm{ac}\sin(2\pi f t_0+\phi)+B_\textrm{dc}\right] +\theta) & A\omega_S\cos(\omega_S \left[B_\textrm{ac}\sin(2\pi f t_0+\phi)+B_\textrm{dc}\right] +\theta) \\
	\vdots & \vdots \\
	A\omega_S\sin(2\pi f t_{N-1}+\phi)\cos(\omega_S \left[B_\textrm{ac}\sin(2\pi f t_{N-1}+\phi)+B_\textrm{dc}\right] +\theta) & A\omega_S\cos(\omega_S \left[B_\textrm{ac}\sin(2\pi f t_{N-1}+\phi)+B_\textrm{dc}\right] +\theta) \\
	\end{array}}\right],
\end{equation}
and the matrix product with its transpose
\begin{equation}
\label{Equation:JacJac non-linear}
J'\times J=A^2\omega_S^2\left[{\begin{array}{cc}
	\sum_{i=0}^{N-1} \sin[2](2\pi f t_i+\phi)\gamma^2\left(t_i\right) & \sum_{i=0}^{N-1} \sin(2\pi f t_i+\phi)\gamma^2\left(t_i\right) \\
	\sum_{i=0}^{N-1} \sin(2\pi f t_i+\phi)\gamma^2\left(t_i\right) & \sum_{i=0}^{N-1} \gamma^2\left(t_i\right) \\
	\end{array}}\right],
\end{equation}
with
\begin{equation}
\gamma\left(t\right)=\cos(\omega_S \left[B_\textrm{ac}\sin(2\pi f t+\phi)+B_\textrm{dc}\right] +\theta).
\end{equation}
Analogue to Supplemental Eq.~(\ref{Equation:JacJac approximation linear}), the sums can be approximated with the knowledge that the measurement is performed during a single period. For $\gamma$ the sum gives
\begin{equation}
\label{Equation:Jac 2x2 non-linear}
\begin{aligned}
\sum_{i=0}^{N-1} \gamma^2\left(t_i\right) &= \sum_{i=0}^{N-1} \cos[2](\omega_S \left[B_\textrm{ac}\sin(2\pi f t_i+\phi)+B_\textrm{dc}\right] +\theta) \approx fN\int_{0}^{\frac{1}{f}}\cos[2](\omega_S \left[B_\textrm{ac}\sin(2\pi f t+\phi)+B_\textrm{dc}\right] +\theta)dt \\ &= \frac{fN}{2}\int_{0}^{\frac{1}{f}}1+\cos(2\omega_S \left[B_\textrm{ac}\sin(2\pi f t+\phi)+B_\textrm{dc}\right] +2\theta)dt \\ &= \frac{N}{2}+\frac{fN}{2}\int_{0}^{\frac{1}{f}}\cos(2\omega_S \left[B_\textrm{ac}\sin(2\pi f t+\phi)+B_\textrm{dc}\right] +2\theta)dt.
\end{aligned}
\end{equation}
The last integral is not straightforward to calculate. The first term inside the cosine oscillates a single period around $0$ with amplitude $2\omega_SB_\textrm{ac}$. Therefore, if $\theta=\pi/4-\omega_SB_\textrm{dc}$, the argument of the cosine oscillates around $\pi/2$, and the integral gives zero. For now, this $\theta$ is assumed. Using both Supplemental Eq.~(\ref{Equation:JacJac approximation linear}) and Supplemental Eq.~(\ref{Equation:Jac 2x2 non-linear}), the sum for the sine times $\gamma$ is
\begin{equation}
\begin{aligned}
\sum_{i=0}^{N-1} \sin[2](2\pi f t_i+\phi)\gamma^2\left(t_i\right) &\approx \frac{fN}{4}\int_{0}^{\frac{1}{f}}\left[1-\cos(4\pi f t+2\phi)\right]\left[1+\cos(2\omega_S \left[B_\textrm{ac}\sin(2\pi f t+\phi)+B_\textrm{dc}\right] +2\theta)\right]dt \\ &= \frac{N}{4} + \frac{fN}{4}\int_{0}^{\frac{1}{f}}-\cos(4\pi f t+2\phi)+\cos(2\omega_S \left[B_\textrm{ac}\sin(2\pi f t+\phi)+B_\textrm{dc}\right] +2\theta) \\ &- \cos(4\pi f t+2\phi)\cos(2\omega_S \left[B_\textrm{ac}\sin(2\pi f t+\phi)+B_\textrm{dc}\right] +2\theta)dt.
\end{aligned}
\end{equation}
The first term in the integral gives zero (double period integration), the second term gives zero with the $\theta$ mentioned above, and the third term evaluates to zero as well with the same $\theta$ since this is the multiplication of an odd and an even periodic function with an average of zero. The remaining question is whether the results are limited by the restriction on $\theta$, since the integrals oscillate around the calculated values. They do not, in similar fashion as in~\cite{Herbschleb2021}, by combining two measurements (for example by alternating between these for each data point) with their $\theta$ $\pi/2$ apart, which is possible by changing the phase of the last MW pulse, the combined result is constant.

Thus, the matrix product and resulting covariance matrix can be approximated as
\begin{equation}
\label{Equation:JacJac non-linear period}
J'\times J\approx A^2\omega_S^2\left[{\begin{array}{cc}
	\frac{N}{4} & 0 \\
	0 & \frac{N}{2} \\
	\end{array}}\right] \Rightarrow 
C=\left(J'\times J\right)^{-1}\approx\frac{1}{A^2\omega_S^2} \left[{\begin{array}{cc}
	\frac{4}{N} & 0 \\
	0 & \frac{2}{N} \\
	\end{array}}\right].
\end{equation}
Therefore, the sensitivities for the non-linear case are $\sqrt{2}$ worse than for the linear case, which is generally the trade-off for increasing the dynamic range~\cite{Herbschleb2021}
\begin{equation}
\begin{aligned}
\eta_\textrm{dc, non-linear} &= \sqrt{2}\eta_\textrm{dc, linear} \\
\eta_\textrm{ac, non-linear} &= \sqrt{2}\eta_\textrm{ac, linear}.
\end{aligned}
\end{equation}

\newpage

\section{Line widths}
\label{Supp:LW}

Several results for repeating the measurement of NMR signals of water are given in Supplemental Fig.~\ref{SFigure:line widths}. These experiments are equivalent to the one performed for main text Fig.~5(a), but a windowed (Hann) Fourier transform is plotted instead to look at the line widths. The Lorentzian $L$ fitted to the data is
\begin{equation}
	\label{Equation:Lorentzian}
	L=a\frac{\left(\frac{\Gamma}{2}\right)^2}{\left(f-f_0\right)^2 + \left(\frac{\Gamma}{2}\right)^2}+b,
\end{equation}
with $\Gamma$ the line width, $f$ the frequency, $f_0$ the frequency of the peak, $a$ the value at $f_0$, though excluding the offset $b$. When fitting the frequency spectra, the maximum value for $a$ is limited to $1.2$ times the maximum data point. This is to ensure the line width does not become unreasonably narrow, as freely fitting a Lorentzian on a peak with just a few (noisy) data points allows the maximum to peak highly above the maximum data point, which decreases the line width considerably.

The fits indicate that the line width for this sample is about $1$$-$$2$~Hz. The line width of the measurement of main text Fig.~5(a) is $1.6$~Hz, slightly above the average here. The line width is limited by the coherence time of the water sample, which is about $0.2$~s in this case.

\begin{figure}[htbp]
	\includegraphics{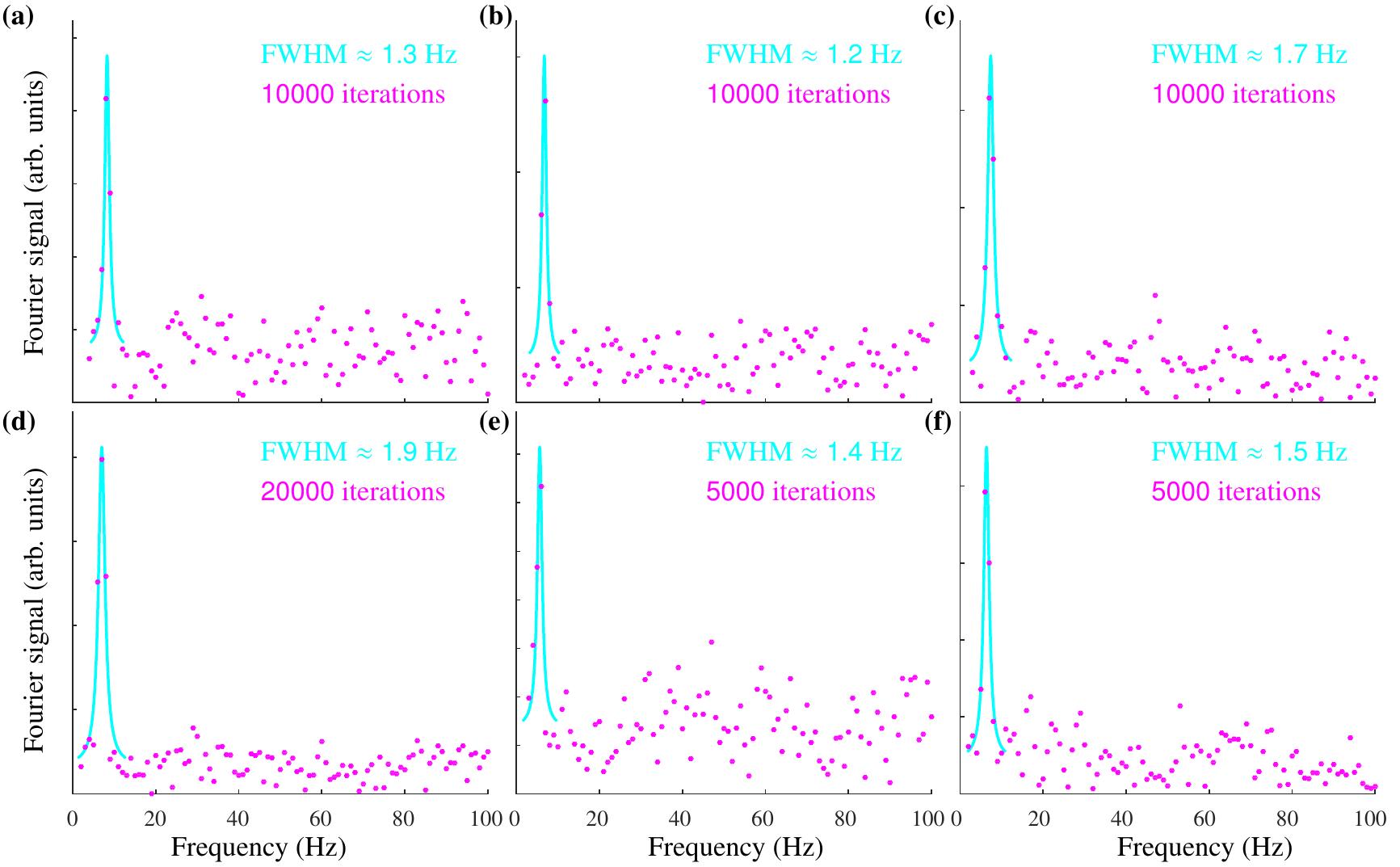}
	\caption{Low frequency regime of Hann-windowed Fourier transforms of NMR signals for a water sample with a sequence length of $1$~s, which data are similar to Fig.~5(a) in the main text. (a) Sequence repeated $10000$ times, line width is $1.3_{-0.1}^{+0.2}$~Hz. (b) $10000$ times, line width $1.2_{-0.1}^{+0.1}$~Hz. (c) $10000$ times, line width $1.7_{-0.1}^{+0.1}$~Hz. (d) $20000$ times, line width $1.9_{-0.1}^{+0.1}$~Hz. (e) $5000$ times, line width $1.4_{-0.2}^{+0.2}$~Hz. (f) $5000$ times, line width $1.5_{-0.1}^{+0.1}$~Hz.}
	\label{SFigure:line widths}
\end{figure}

\newpage

%
